\documentclass[twocolumn,superscriptaddress,amssymb,amsmath,nobibnotes,aps,prl]{revtex4-2}

\pdfoutput=1

\usepackage{color}
\usepackage{graphicx}
\usepackage{amsmath}
\usepackage{mathtools}
\usepackage{bbold}
\usepackage{stmaryrd}
\usepackage{amsmath}
\usepackage{amsfonts}          
\usepackage{amssymb}           
\usepackage{graphicx}  
\usepackage{mathrsfs}
\usepackage{float}
\usepackage{url}
\usepackage{setspace}
\usepackage{braket}
\usepackage{cancel}
\usepackage{csquotes}
\usepackage{dsfont}
\usepackage{physics}
\usepackage{algorithm}
\usepackage{algorithmic}
\usepackage[hypertexnames=false]{hyperref}
\usepackage{comment}

\DeclareMathAlphabet{\mathpzc}{OT1}{pzc}{m}{it}

\newcommand{\ijk}[3]{\alpha_{#1#2#3}}

\newcommand{\myroundedbrackets}[1]{\left(#1\right)}

\newcommand{\inner}[2]{\langle#1,#2\rangle}
\newcommand{\ijkvec}[3]{\alpha_{\vec{#1}\vec{#2}\vec{#3}}}

\newcommand{\oforder}[1]{\mathcal{O}(#1)}
\newcommand{\Tm}[2]{{#1}_{\Vec{#2}}}

\begin{document}

\title{Quantum Complexity as Hydrodynamics}

\author{Pablo Basteiro}
\affiliation{Institut f\"ur Theoretische Physik und Astrophysik and \\ W\"urzburg-Dresden Cluster of Excellence ct.qmat,\\ Julius-Maximilians-Universit\"at W\"urzburg, \\Am Hubland, 97074 W\"urzburg, Germany  }
\author{Johanna Erdmenger}
\affiliation{Institut f\"ur Theoretische Physik und Astrophysik and \\ W\"urzburg-Dresden Cluster of Excellence ct.qmat,\\ Julius-Maximilians-Universit\"at W\"urzburg, \\Am Hubland, 97074 W\"urzburg, Germany  }
\author{Pascal Fries}
\affiliation{Institut f\"ur Theoretische Physik und Astrophysik and \\ W\"urzburg-Dresden Cluster of Excellence ct.qmat,\\ Julius-Maximilians-Universit\"at W\"urzburg, \\Am Hubland, 97074 W\"urzburg, Germany  }
\author{\\ Florian Goth}
\affiliation{Institut f\"ur Theoretische Physik und Astrophysik and \\ W\"urzburg-Dresden Cluster of Excellence ct.qmat,\\ Julius-Maximilians-Universit\"at W\"urzburg, \\Am Hubland, 97074 W\"urzburg, Germany  }
\author{Ioannis Matthaiakakis}
\affiliation{Institut f\"ur Theoretische Physik und Astrophysik and \\ W\"urzburg-Dresden Cluster of Excellence ct.qmat,\\ Julius-Maximilians-Universit\"at W\"urzburg, \\Am Hubland, 97074 W\"urzburg, Germany  }
\affiliation{Dipartimento di Fisica, Universit\`a di Genova and \\ I.N.F.N. - Sezione di Genova, \\ via Dodecaneso 33, I-16146, Genova, Italy}
\author{Ren\'e Meyer}
\affiliation{Institut f\"ur Theoretische Physik und Astrophysik and \\ W\"urzburg-Dresden Cluster of Excellence ct.qmat,\\ Julius-Maximilians-Universit\"at W\"urzburg, \\Am Hubland, 97074 W\"urzburg, Germany  }
\email{Corresponding author: rene.meyer@physik.uni-wuerzburg.de}

\date{\today}

\begin{abstract}
As a new step towards defining complexity for quantum field theories, we map Nielsen operator complexity for $SU(N)$ gates to two-dimensional hydrodynamics. We develop a tractable large $N$ limit that leads to regular geometries on the manifold of unitaries as $N$ is taken to infinity. To achieve this, we introduce a basis of non-commutative plane waves for the $\mathfrak{su}(N)$ algebra and define a metric with polynomial penalty factors. Through the Euler-Arnold approach we identify incompressible inviscid hydrodynamics on the two-torus as a novel effective theory of large-qudit operator complexity. For large $N$, our cost function captures two essential properties of holographic complexity measures: ergodicity and conjugate points.
\end{abstract}
 
\maketitle

\textit{1. Introduction.--} Quantum computational complexity \cite{watrous2008quantum}, referred to as complexity hereafter, quantifies the number of simple gates required to synthesize a given unitary operation in quantum computing. In recent years, the geometric approach to complexity of Nielsen et~al.\,\cite{NielsenLowerBounds,NielsenQCasGeometry,NielsenGeometryOfQC} has proven immensely useful for investigating the complexity of $n$-qubit systems, independently of the particular state of the system. In its original manifestation, this geometric framework relied on the $SU(2^n)$ manifold of unitaries acting on $n$-qubits. In this Letter, we consider a generalization of Nielsen's approach that incorporates quantum circuits acting on $N$-level systems, i.e. qudits of dimension $N$. For qudit systems, the unitaries of interest belong to $SU(N)$. The key ingredient in Nielsen's approach is the choice of metric on this manifold, assigning penalty factors to unitaries departing from the identity operator $I$. The complexity of $U\in SU(N)$ is then identified with the length of the minimal geodesic connecting $I$ and $U$. These unitaries are generated by a control Hamiltonian $\mathcal{H}$ tangent to $SU(N)$. The corresponding group algebra characterizes fully these Hamiltonians and their geodesics, through the Euler-Arnold method \cite{PhysRevLett.122.231302,erdmenger2020complexity,Flory:2020eot,Flory:2020dja}.  

Recent progress on complexity measures via the AdS/CFT correspondence (or holography) \cite{Maldacena:1997re,Witten:1998qj,Gubser:1998bc} further motivates the present work. In general, holographic complexity measures should be ergodic and exhibit conjugate points: Ergodicity ensures all points on the group manifold can be reached in finite time, thus implying a linear \footnote{See \cite{Haferkamp:2021uxo} for a recent proof of this linear growth of complexity for Haar-random circuits.} growth of complexity with time \cite{Brown:2016wib,Brown:2017jil}. In contrast, conjugate points, i.e. the meeting points of equal-length geodesics, provide bounds on this growth \cite{NielsenGeometryOfQC}. This conjecture is supported by work on the curvatures of complexity measures in holographic CFTs \citep{Flory:2020dja,Flory:2020eot}.

Based on Nielsen's approach, different definitions for complexity both for discrete systems \cite{Brown:2016wib,Brown:2017jil,Brown:2019whu,Balasubramanian:2019wgd,Auzzi:2020idm,Balasubramanian:2021mxo} and quantum field theories \cite{Magan:2018nmu,Jefferson_2017,Chapman:2017rqy,Khan:2018rzm,Hackl_2018,Chapman_2019,erdmenger2020complexity,Flory:2020dja,Flory:2020eot} have been investigated. However, how these notions of complexity are related in the limit of infinite Hilbert space dimensions is an open question.
This limit is in general not well-defined, because desirable features of quantum circuits, such as $k$-locality \cite{Brown:2017jil,Balasubramanian:2019wgd,Balasubramanian:2021mxo,Auzzi:2020idm,Wu:2021pzg}, require penalty factors typically scaling exponentially with the system size for every direction on the manifold \cite{NielsenLowerBounds}. In the $N\rightarrow \infty$ limit, this leads to singular geometries on the manifold of unitaries, impeding the definition of complexity as geodesic length. 

In this Letter, we show how a judicious choice of basis and metric for the $\mathfrak{su}(N)$ algebra with \textit{polynomial} penalty factors leads to well-defined non-singular geometry for $SU(N\rightarrow\infty)$ \footnote{The large $N$ limit we consider is similar to the vector large $N$ limit of $O(N)$ models of quantum field theories, where fields transform in the fundamental representation of the symmetry group, and the number of degrees of freedom $N$ is taken to infinity \cite{Klebanov:2018fzb}.}. We show that, on the $SU(N)$ manifold at infinite $N$ and at low energies, Nielsen complexity can be equivalently evaluated on the manifold of volume-preserving diffeomorphisms SDiff($\mathds{T}^2$) of the torus. We show that the Euler-Arnold equation on the resulting manifold coincides with the Euler equation of a two-dimensional ideal fluid \footnote{We define an ideal fluid as being incompressible and inviscid.}. This permits identifying control Hamiltonians with diffeomorphism generators in two-dimensional hydrodynamics. Moreover, it suggests a natural cost function, based on the two-dimensional Laplacian, with a smooth dependence on $N$. This smooth dependence suggests the prevalence of particular characteristics of the hydrodynamic theory even at finite $N$. In particular, two-dimensional ideal hydrodynamics is classically chaotic due to the hyperbolic geometry of its phase space \cite{arnold2008topological}. We quantify this instability at finite, large $N$ by numerically computing the sectional and Ricci curvatures of $SU(N)$. 

Finally, we find that our formulation of complexity  exhibits both ergodicity and conjugate points, whose presence is necessary for a proper holographic complexity measure (although see \cite{Note2}). In this way, our results constitute a new step towards understanding quantum complexity in QFT's with a holographic dual.

\hypertarget{Section2}{\textit{2. The algebra of $N$-level qudits.--}} At the heart of our construction lies a new and non-trivial choice of anti-Hermitian generators for the $\mathfrak{su}(N)$ Lie algebra. We employ known results for $\mathfrak{su}(N)$ to investigate the large $N$ limit of our basis and explain the subtleties it carries. The corresponding structure constants determine the Riemannian curvature of $SU(N)$, which we compute in section \hyperlink{Section5}{5}. We outline our construction below \footnote{See Supplemental Material [URL inserted by the publisher] for more details.}. 

We first introduce the $N\times N$ \enquote{shift} matrix $h_{ij}=\delta_{i+1,j}$ and \enquote{clock} matrix $g_{ij}=\omega^j\delta_{i,j}$, with $\omega=\exp\frac{2\pi i}{N}$ a primitive $N^{\textrm{th}}$ root of unity, and $i,j=0,\dots,N-1\;\textrm{mod}\,N$. These matrices commute up to a phase, i.e. $hg=\omega gh$. Then, following \cite{FAIRLIE1989203,FAIRLIE1989101,Fairlietrigonometric,Patera}, we define a basis of unitary, but not necessarily anti-Hermitian, matrices $J_{\vec{m}}=\omega^{\frac{m_1m_2}{2}}g^{m_1}h^{m_2}$, indexed by a two-vector $\vec{m}=(m_1,m_2)$ on the $\mathds{Z}^2$-lattice. These can be thought of as a non-commutative version of plane waves, with the vector index $\vec{m}$ playing the role of the wave vector, and $h$ and $g$ the momentum and position modes, respectively \cite{Fairlietrigonometric}. Their commutator is given by
\begin{equation}
    [J_{\vec{m}},J_{\vec{n}}]= -2i\sin\left(\frac{\pi}{N}(\vec{m}\times \vec{n})\right)\,J_{\vec{m}+\vec{n}}\,,
    \label{AlgebraForTheJs}
\end{equation}
where $\vec{m}\times\vec{n}\equiv m_1n_2-m_2n_1$. In \cite{Fairlietrigonometric} it was shown that the algebra \eqref{AlgebraForTheJs} is, in the $N\rightarrow \infty$ limit, isomorphic to the algebra SVect($\mathds{T}^2$) of the group SDiff($\mathds{T}^2$) of volume-preserving diffeomorphisms on the standard two-torus $\mathds{T}^2$. To see this, note that SVect($\mathds{T}^2$) admits a symplectic structure in terms of divergence-free vector fields which, in two-dimensions, are Hamiltonian vector fields $X_f$. The $X_f$ are uniquely determined by their associated \textit{stream function} $f$ \footnote{We show that the stream functions defined here are in one-to-one correspondence with the stream functions of hydrodynamics in section \unexpanded{\hyperlink{Section3}{3}}, see also \cite{Note4}.}
, which can be expanded in plane waves on $\mathds{T}^2$ as $f_{\vec{m}}\propto\textrm{exp}(i(m_1x+m_2p))$ \cite{arnold2008topological}. The isomorphism then between SVect($\mathds{T}^2$) and $\mathfrak{su}(N)$ in the large $N$ limit is obtained by expanding the sine in \eqref{AlgebraForTheJs} to first order in $1/N$ and identifying 
\begin{equation}
    J_{\vec{m}}\overset{N\rightarrow\infty}{\longrightarrow} \frac{2\pi}{iN}X_{\vec{m}}\,.
    \label{AlgebraIsomoprhism}
\end{equation} 
The isomorphism \eqref{AlgebraIsomoprhism} thus relates Hamiltonian vector fields (i.e. elements of SVect($\mathds{T}^2$)) with the basis elements of $\mathfrak{su}(N)$ given by \eqref{AlgebraForTheJs}. A detail not addressed in \cite{FAIRLIE1989203,FAIRLIE1989101,Fairlietrigonometric,Patera}, but already mentioned in \cite{PhysRevA.46.6417}, is that the Taylor expansion truncation is invalid for several classes of vectors $\vec{m}, \vec{n}$; There are vector pairs defined for all $N$, e.g. $\vec{m}=(\frac{N-1}{2},0)$ and $\vec{n}=(0,\frac{N-1}{2})$, for which the cross product is of order $\oforder{N^2}$ or $\oforder{N}$. We must restrict the isomorphism to only the pairs of $\oforder{1}$. These turn out to be precisely the low-momentum modes relevant for hydrodynamics, see \hyperlink{Section4}{Sec. 4} and \cite{Note4} for more details.

We now proceed with the definition of anti-Hermitian basis elements, capable of constructing $SU(N)$ operators through exponentiation, as required by Nielsen's approach. To this end, we introduce for each $\vec{m}$,
\begin{equation}
    C_{\vec{m}}\equiv i(J_{\vec{m}}+J^{\dagger}_{\vec{m}})\,,\quad
    S_{\vec{m}}\equiv (J_{\vec{m}}-J^{\dagger}_{\vec{m}})\,.
    \label{DefinitionOfTheCsAndSS}
\end{equation}
The generators \eqref{DefinitionOfTheCsAndSS} obey commutation relations inherited from \eqref{AlgebraForTheJs} \cite{Note4}. The structure constants thus obtained are more involved than those in \eqref{AlgebraForTheJs} due to the overcompleteness of \eqref{DefinitionOfTheCsAndSS}. However, we can make this basis complete via modularity symmetries and linear dependences enjoyed by the generators \cite{Note4}. Most importantly, due to linearity, the isomorphism \eqref{AlgebraIsomoprhism} carries over to this basis, which hence exhibits a well-defined large $N$ limit. We exploit this in our curvature computations in section \hyperlink{Section5}{5}.

\hypertarget{Section3}{\textit{3. Euler-Arnold framework.--}} Nielsen's approach identifies complexity with the length of the minimal geodesic $U(s)$, with $s$ parametrizing the position along the trajectory \footnote{In the context of holography, one is interested in target unitaries describing the time evolution of the system under a physical Hamiltonian $H$ up to a given time $t$, i.e. $U=e^{Ht}$. The parameter $s$ in the \textit{control} Hamiltonian $\mathcal{H}$ should not be confused with the time $t$ of the \textit{physical} Hamiltonian $H$, as these are in general not equivalent. We assume the physical Hamiltonian implements all-to-all level transitions within the qudit \cite{Note4}.}, connecting the identity element with the desired $U$ on the manifold of unitaries. Each geodesic $U(s)$ is generated by an (anti-Hermitian) \textit{control Hamiltonian} $\mathcal{H}(s)$ via the Schrödinger equation $\frac{dU}{ds}=\mathcal{H}(s)U(s)$. 
The Euler-Arnold formalism \cite{AIF_1966__16_1_319_0,arnold2008topological} exploits the group structure by identifying $\mathcal{H}$ with a Lie algebra element $\mathcal{H}(s)=U^{-1}(s)\Dot{U}(s)\in\mathfrak{su}(N)$, i.e. with the pullback of the vector $\dot{U}(s)$ onto the tangent space at the identity. The time-evolution of $\mathcal{H}(s)$ within this approach is then given by the \textit{Euler-Arnold} (EA) equation

\begin{equation}
	\dot{\mathcal{H}}=\kappa(\mathcal{H},\mathcal{H})\,,
	\label{EulerArnoldEquation}
\end{equation}

where $\kappa$ is a quadratic bilinear two-form defined via  $\left<[X,Y],Z\right>=\left<\kappa(Z,X),Y\right>$, for $X,Y,Z\in\mathfrak{su}(N)$ and $\left<\cdot,\cdot\right>$ the Lie algebra inner product \cite{AIF_1966__16_1_319_0}. Distinct $\kappa$-forms are induced by different inner products on the algebra. This is equivalent to choosing a metric on $SU(N)$, and hence a cost function in Nielsen's setup. Using the EA equation is advantageous, since it can be easier to solve \eqref{EulerArnoldEquation} than to compute the nested commutators appearing in a solution of the Schrödinger equation \cite{Magan:2018nmu}. In fact, we explain in the next section how the EA equation drastically simplifies in the large $N$ limit, which allows for the direct calculation of the control Hamiltonian.

\hypertarget{Section4}{\textit{4. Inner product and penalty factors.--}} We now show how at large $N$, $\mathfrak{su}(N)$ leads within the EA framework  to the ideal-fluid Euler equation. Solutions to this equation are control Hamiltonians in the sense of Nielsen, which allows for a hydrodynamic interpretation of the standard computation of Nielsen complexity \cite{Note4}. We also discuss in detail how this suggests a natural extrapolation of the hydrodynamic cost function to finite $N$.

In the large $N$ limit and for low-energy $\oforder{1}$ generators, $SU(N)$ is identified via \eqref{AlgebraIsomoprhism} with the manifold of volume-preserving diffeomorphisms SDiff($\mathds{T}^2$) \footnote{The restriction to $\oforder{1}$ generators is justified by our choice of penalty factors as we elucidate in the following.}. 
We consider the standard inner product on its algebra SVect($\mathds{T}^2)$, given by the $L^2$-inner product between Hamiltonian vector fields. This can be rewritten in terms of the Laplacian acting on stream functions as
\begin{equation}
    \langle X_f,X_g\rangle=\int_{\omega} X_f\cdot X_g=-\int_{\omega} f\Delta g\,,
    \label{L2Product}
\end{equation}
 with $\omega$ the symplectic form on $\mathds{T}^2$ \cite{Note4}. For $f = g$, this is (twice) the kinetic energy of a flow. 
The inner-product \eqref{L2Product} induces a metric on SDiff($\mathds{T}^2$) which defines the length, and thus the cost, of geodesics. Consequently, it defines a $\kappa$ form given by $\kappa(f,f) =-\Delta^{-1}\{f,\Delta f\}$, with $\{\cdot,\cdot\}$ the usual Poisson bracket \cite{Note4}. This $\kappa$ form leads to the following EA equation for the control Hamiltonian $\mathcal{H}$ 
\begin{equation}
    \Delta \dot{\mathpzc{h}}=-\{\mathpzc{h},\Delta \mathpzc{h}\}\,,
    \label{EAEquationHydro}
\end{equation} 
with $\mathpzc{h}$ the \enquote{control stream function} associated to $\mathcal{H}$ via the symplectic form \cite{Note4}.
Equation \eqref{EAEquationHydro} constitutes a main result of this work. Considering the large $N$ limit of the $\mathfrak{su}(N)$ algebra, we obtain the stream function form of the Euler equation for a $(2+1)$-dimensional ideal fluid \cite{arnold2008topological}.

We find that the Nielsen complexity of a large-qudit unitary can be evaluated via this effective hydrodynamic theory, for which the control Hamiltonian can be straightforwardly computed \eqref{EAEquationHydro}.  Moreover, the computation of Nielsen complexity can be recast in the hydrodynamic setting: The Schrödinger equation defining the target unitary $U$ in terms of the control Hamiltonian, is now the equation relating the Eulerian and Lagrangian frames of reference of the fluid $\frac{d f}{ds}=\mathcal{H}(s,f(s))$ \cite{goldstein2002classical}. 
$\mathcal{H}(s)$ is the control Hamiltonian obtained from the solution $\mathpzc{h}$ of \eqref{EAEquationHydro} and $f$ the corresponding diffeomorphism. Imposing the boundary conditions $f(0)=\mathbf{id}$, the identity map, and $f(1)=f_{\textrm{target}}$, the target diffeomorphism \footnote{Here, $f_{\textrm{target}}$ can be also taken to be time evolution of the fluid, assuming the physical Hamiltonian obeys similar properties as the aforementioned qudit Hamiltonian \cite{Note4}.}, yields initial velocities $v^{\vec{m}}\equiv v^{\vec{m}}(s,f_{\textrm{target}})$ \footnote{See eq.~(S20) in \cite{Note4}.} for the geodesic as functions of $f_{\textrm{target}}$. These are inserted into the length functional $\ell=\int_0^1\,ds\sqrt{\tilde{G}_{\vec{m}\vec{n}}v^{\vec{m}}v^{\vec{n}}}$. Here, $\tilde{G}_{\vec{m}\vec{n}}$ is the metric induced by the inner product \eqref{L2Product}. Minimizing this functional over all solutions \eqref{EAEquationHydro} yields the complexity of $f_{\textrm{target}}$, ${\cal C}(f_{\rm target})$. In summary, the Nielsen complexity of large-qudit unitaries is given by the length of the minimal geodesic, generated by a solution to the EA equation \eqref{EAEquationHydro}, that connects the Lie algebra SVect($\mathds{T}^2)$ with the desired target element of SDiff($\mathds{T}^2$).  Our construction, hence, provides a smooth geometry at large $N$, the manifold SDiff($\mathds{T}^2$), on which complexity can be calculated. Thus, we avoid the singular geometries encountered in previous manifestations of large $N$ complexity models. Additionally, SDiff($\mathds{T}^2$) has a clear physical interpretation as the phase space of a well-known theory, namely two-dimensional hydrodynamics.

\begin{figure}[t]
    \centering
    \includegraphics[width=\columnwidth]{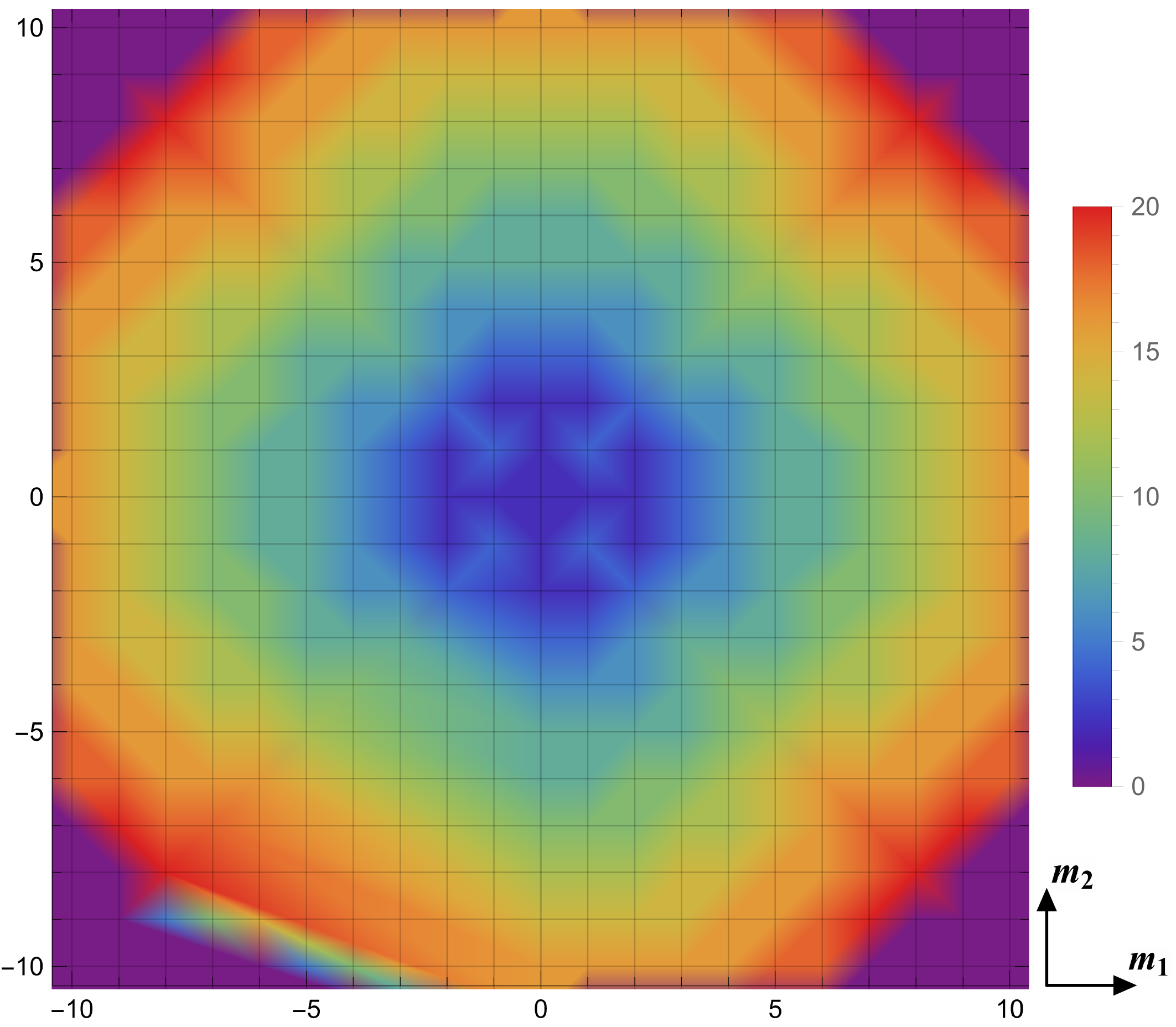}
    \caption{Color density plot of the critical value $N_c$ at which the normalized Ricci curvature of a given direction $\vec{m}$ in $\mathfrak{su}(N)$ becomes negative over the $\mathds{Z}^2$ lattice, spanned by the vector components $m_1,m_2$. The interpolation between the integer points of the lattice is there to guide the eye. The color flare at the lower left corner is an artifact of this interpolation. }
    \label{MainPlot}
\end{figure}

Motivated by our construction for complexity at $N\rightarrow\infty$, and exploiting the generator isomorphism \eqref{AlgebraIsomoprhism}, we now formulate new results for the complexity geometry at finite $N$ by ensuring a smooth transition between the finite- and infinite-dimensional setups. We adapt the inner product \eqref{L2Product} at $N\rightarrow \infty$ to finite $N$ by defining the action of the Laplacian on non-commutative waves as $\Delta J_{\vec{m}}=-m^2J_{\vec{m}}$. This choice transitions smoothly to infinite $N$, where the action of the standard Laplacian on plane waves $f_{\vec{k}}$ is $\Delta\,f_{\vec{k}}=(\partial^2_x+\partial^2_p)\,f_{\vec{k}}=-k^2f_{\vec{k}}$ \footnote{Our definition of the discrete Laplacian is the canonical one but it is non-unique, see e.g. \cite{PhysRevA.46.6417}.}. We define an inner product on $\mathfrak{su}(N)$ for finite $N$ as
\begin{equation}
    \langle \mathcal{T}_{\vec{m}},\mathcal{T}_{\vec{n}}\rangle := -\frac{1}{2N}\Tr(\mathcal{T}_{\vec{m}}\Delta \mathcal{T}^{\dagger}_{\vec{n}})\,,
    \label{InnerProduct}
\end{equation}
with $\mathcal{T}\in\{C,S\}$. By means of group translation, this inner product induces a right-invariant metric $G_{\vec{m}\vec{n}}$ on $SU(N)$. Its components are the penalty factors for different directions on the tangent space, given by the eigenvalues of the Laplacian acting on the generators, i.e. $G_{\vec{m}\vec{n}}=m^2 \delta_{\vec{m}\vec{n}}$, with $m=\abs{\vec{m}}$. These penalties render the metric homogeneous but \textit{not} isotropic, since not all directions get penalized equally. Equal penalty factors are assigned only to those vectors related by parity or conjugation e.g. $\vec{m}=(1,2)$, $\vec{n}=(2,1)$ and $\vec{l}=(-2,-1)$. This reflects the Hamiltonian structure of the problem, with the Laplacian being invariant under symplectic transformations, and is visually manifest in our curvature results shown in Fig.~\ref{MainPlot}. Due to this symmetry, every direction on the Lie algebra gets assigned a different penalty with at most eight-fold degeneracy, yielding a maximally anisotropic metric for the manifold of unitaries. 
This is an essential property, since anisotropy leads to negative curvature on the manifold and negative curvature is a strong indicator of ergodic geodesic flow \cite{Brown:2019whu}. In terms of $N$-level qudits, our choice of metric ensures high-energy excitations with large wave vector receive larger penalty factors. These high-energy sectors effectively decouple in the strict large $N$ limit.

Our choice of penalty factors is fundamentally different from the majority of previous work on the subject, e.g. \cite{NielsenGeometryOfQC,Brown:2017jil,Balasubramanian:2019wgd,Auzzi:2020idm}. In particular, the penalty factors in these setups grow exponentially $p\sim \alpha^k\sim e^{k\ln{\alpha}}$ instead of polynomially \footnote{See however \cite{Magan:2018nmu} where polynomial penalties were first suggested.}. Here, $k$ is the Pauli weight of the many-qubit gate in the Pauli basis \footnote{The Pauli basis is used to decompose many-qubit gates into tensor products of the standard Pauli matrices and the identity. A gate is of weight $k$  (or \textit{$k$-local}) if its tensor product contains up to $k$ Pauli matrices.}. Although exponential penalty factors are well motivated from the point of view of local quantum operations, they typically lead to singular geometries in the $N=2^n\rightarrow \infty$ limit \cite{Auzzi:2020idm,Balasubramanian:2019wgd,Balasubramanian:2021mxo}. Instead, our penalty factors remain finite in the large $N$ limit, by transitioning to the position and momentum modes of Hamiltonian vector fields on $\mathds{T}^2$ via \eqref{AlgebraIsomoprhism}. This identification with the hydrodynamical phase-space is naturally restricted to the low-energy sector of $\oforder{1}$ vectors, the so-called \textit{admissible} directions. Thus, there are no infinitely penalized directions on SDiff($\mathds{T}^2$), resulting in a regular geometry. 
The remaining directions of the large $SU(N\rightarrow\infty)$ manifold containing the high-energy sectors with $\oforder{N}$ vectors are \textit{inadmissible} and effectively decouple from the geometry since they are assigned penalty factors that are \textit{at least} infinite. This situation is captured in the framework of sub-Riemannian geometry \cite{agrachev2019}, also recently mentioned in the context of complexity in \cite{Brown:2021rmz}. A fundamental theorem due to Chow and Rashevskii \cite{Rashevskii1938,Chow1940} asserts that geodesics can still reach every point by only accessing admissible directions \cite{KhajehSalehani2014ControllabilityOI}\footnote{The applicability of this theorem to our setup is discussed in detail in the SM.}. That is, trajectories from the hydrodynamic phase space can still reach every large qudit unitary. Since it is infinitely expensive to move in inadmissible directions, the hydrodynamic trajectories have an overall smaller cost, i.e. smaller complexity, cf.~\cite{Note4}.

\hypertarget{Section5}{\textit{5. Average Ricci curvature--}}
Curvature computations in recent literature regarding Nielsen complexity geometries \cite{Brown:2017jil,Brown:2019whu,Balasubramanian:2019wgd,Auzzi:2020idm,Balasubramanian:2021mxo,Wu:2021pzg} typically focus on the sectional curvatures of the manifold of unitaries. The sign of the sectional curvatures is an indicator for convergence (positive sign) or divergence (negative sign) of nearby geodesics  \cite{Arnold2014}. However, the stability of a geodesic and, hence, the emergence of ergodic behaviour does not only depend on the sign of the sectional curvature in the direction parallel to its velocity, but rather on the sign of the sectional curvatures of all two-planes containing its velocity vector \cite{arnold2008topological}. For this reason, we believe that a more telling quantity to describe the stability of a geodesic with velocity vector $v$ is given by the \textit{normalized Ricci curvature} \cite{Lukatskii1981},
\begin{equation}
    Ric(v)=\lim_{N\rightarrow \infty}\frac{1}{N^2-2}\sum_{\vec{m}}K(v,\mathcal{T}_{\vec{m}})\,,
    \label{NormalizedRicciCurvature}
\end{equation}
with $K$ the sectional curvature tensor and $\vec{m}$ running over the algebra directions. Eq.~\eqref{NormalizedRicciCurvature} can be thought of as an average sectional curvature across an orthonormal basis for the tangent space and is well-defined as $N\rightarrow\infty$. Importantly, $Ric(v)\leq 0$ for SDiff($\mathds{T}^2$) \cite{Lukatskii1981}, which is the quantitative reason for the chaotic behavior of two-dimensional hydrodynamics. The smooth limit of our $\mathfrak{su}(N)$ basis for large $N$, given by SVect$(\mathds{T}^2)$, indicates we can compute Ricci curvatures of $SU(N)$ at large $N$ and compare to the hydrodynamic result. We evaluate the Ricci curvature for every direction $\vec{m}$ in $\mathfrak{su}(N)$ for odd values of $N\in[3,39]$ by first calculating the corresponding sectional curvatures \cite{Note4}. Note that, since the dimensionality of the tangent space grows with $N$, a given velocity $\vec{m}$ can be defined only after it appears within the distribution of directions at $N = N_0(\vec{m})$. Its corresponding Ricci curvature $Ric(\vec{m})$ is thus defined only after $ N = N_0(\vec{m})$ and will continue to change with $N$ as more and more directions contribute to the average in \eqref{NormalizedRicciCurvature}. Our numerical data shows that Ricci curvatures of newly introduced directions at a given $N= N_0$ are always positive, but all eventually turn negative at some \textit{critical value} $N = N_c(\vec{m})$. The resulting data for $N_c$ as a function of the direction $\vec{m} \in\mathfrak{su}(N)$ is shown in Fig.~\ref{MainPlot} and constitutes a second main result of our work. We interpret this figure and its extrapolation at large $N$ as a visual definition of the $\oforder{1}$ sub-sector (the blue region) from which the hydrodynamic theory emerges in the strict large $N$ limit \cite{Note4}. 

Our results have the following implications for the complexity geometry of $SU(N)$ at large $N$. The large $N$ geometry of the low-energy sector has negative Ricci curvature, thus numerically confirming previous mathematical results \cite{Lukatskii1981}, as well as indicating emergent chaotic behaviour \footnote{A similar behavior of negative curvature only in a subsector of the manifold of unitaries, built out of the Pauli basis, was found in the context of operator size complexity in \cite{Wu:2021pzg}.}. This implies geodesics can reach every point of the manifold in finite time, resulting in an ergodic geodesic flow. Therefore, our canonical cost function has a property characteristic of any proper holographic complexity measure as conjectured by \cite{Brown:2016wib,Brown:2017jil}. 

Furthermore, the numerically-evaluated distribution of sectional curvatures of our model always contains positive sectional curvatures at large $N$ \cite{Note4}. The presence of positively sectional curvatures is also a necessary property of our cost function from two points of view: First, while a strictly negative geometry is indeed ergodic  \cite{MR0224110,Auzzi:2020idm}, complexity metrics on Lie groups without positive curvatures are necessarily flat \cite{Milnor,Balasubramanian:2021mxo}. Our choice of metric thus combines the negative average Ricci curvature beneficial for ergodicity with the necessity of having positively curved directions.
Second, strictly negative geometries lack an important feature of geometric complexity, namely conjugate points \cite{bishop2013riemannian}. These are points on a manifold where a geodesic ceases to be globally minimizing, e.g. antipodal points on the sphere. Conjugate points seem to be a necessary feature of complexity geometries, since they forbid complexity from exhibiting an unbounded linear growth with time \cite{NielsenGeometryOfQC,Balasubramanian:2019wgd,Auzzi:2020idm,Balasubramanian:2021mxo}. It is well-known that  SDiff($\mathds{T}^2$), our effective geometry at $N\rightarrow\infty$, indeed exhibits conjugate points \cite{arnold2008topological,10.2307/2161966}.
All in all, our results indicate that even though the large $N$ limit considered here is more similar to the vector large $N$ limit than to the matrix large $N$ limit relevant in holography \cite{Klebanov:2018fzb}, our setup still exhibits desirable properties of a holographic complexity measure for large enough values of $N$.

Finally, our results for the sectional curvatures \cite{Note4} indicate the existence of a universality class of $SU(N)$ metrics, as defined in \cite{Brown:2021rmz}, indexed by $N^2$. These $SU(N)$ metrics are conjectured to be equivalent, i.e. leading to the same complexity, at late geodesic times. We infer from this conjecture that the complexity of $SU(N)$ scales at large $N$ and at large geodesic distances as ${\cal C}(SU(N)) \simeq {\cal C}({\rm SDiff}(\mathds{T}^2)) + \oforder{1/N}$. Moreover, this implies a finite critical $N=N_C$ such that ${\cal C}(SU(N_C))$ at short distances equals ${\cal C}({\rm SDiff}(\mathds{T}^2))$ at long distances, even if the manifolds are not isomorphic.

\hypertarget{Section6}{\textit{6. Discussion and Outlook.--}} For the first time, we provide a definition for Nielsen operator complexity of $SU(N)$ with a well-defined large $N$ limit. This is realized by using the ideal hydrodynamics equation as the geodesic equation on the low-momentum sector of $SU(N\rightarrow \infty)\cong \textrm{SDiff(}\mathds{T}^2\textrm{)}$. The natural choice of cost function is the kinetic energy of the fluid, which we derive within the Euler-Arnold approach and adapt for every value of $N$. Our construction provides a simple way of computing the control Hamiltonian, thus simplifying one of the main obstacles in the computation of Nielsen complexity. In particular, our setup allows to reach every point of the large $SU(N\rightarrow
\infty)$ manifold via admissible geodesics within the hydrodynamical phase space, thus drastically simplifying the complexity geometry.

From the perspective of quantum information, we find a basis for qudit unitaries that scales nicely with the qudit size. This scalability allows for the synthesis of large qudit gates as long as they only implement $\oforder{1}$ transitions, with respect to $N$. This corresponds to a locality property of our basis, implying its usefulness for constructing qudit lattices. This is particularly interesting in view of computing complexity of fault-tolerant quantum error-correction qudit architectures \cite{Gottesman:1998se,doi:10.1137/S0097539799359385,PhysRevLett.113.230501}. 

It is possible to include $1/N$ corrections for large, but finite, qudit unitaries in order to confirm the conjectured scaling of ${\cal C}(SU(N))$ with $N$ in terms of ${\cal C}($SDiff($\mathds{T}^2$)). This will also allow to derive the critical $N_C$ at which the complexities for the two manifolds coincide. This approach is closely related to integrable systems in non-commutative geometry \cite{PhysRevA.46.6417,connes1995noncommutative,Hoppe1992,Khesin_2004}, see \cite{Note4} for a first step in this direction.

Finally, our cost function captures two essential properties in view of holographic complexity, ergodicity and conjugate points. Both are consequences of the phase space geometry of hydrodynamics. This suggests that cost functions based on the Laplacian acting on infinitesimal gates are a promising new avenue for describing operator complexity also in holographic CFTs.

\begin{acknowledgments}
{\bf Acknowledgements.} We thank Stefan Waldmann and Knut Hüper for useful discussions. We thank R.~Auzzi and N.~Zenoni for pointing out a sign error in the first version of this paper. P.B., J.E., I.M.~and R.M.~acknowledge support by the Deutsche Forschungsgemeinschaft (DFG, German Research Foundation) under Germany's Excellence Strategy through the W\"urzburg‐Dresden Cluster of Excellence on Complexity and Topology in Quantum Matter ‐ ct.qmat (EXC 2147, project‐id 390858490). J.E., F.G., I.M. and R.M.~furthermore acknowledge financial support through the  Deutsche  Forschungsgemein\-schaft  (DFG,  German  Research  Foundation),  project-id   258499086   -   SFB   1170   ’ToCoTronics. P.F.~was supported by the DFG project HI 744/9-1. I.M.~has been partially supported by the ``Curiosity Driven Grant 2020'' of the University of Genoa and by the INFN Scientific Initiatives SFT: ``Statistical Field Theory, Low-Dimensional Systems, Integrable Models and Applications''. Finally, the authors gratefully acknowledge the computation resources and support provided by the Universität Würzburg IT Center and the German Research Foundation (DFG) through grant No. INST 93/878-1 FUGG. 
\end{acknowledgments}

\clearpage

\begin{widetext}

\section{Supplemental Material}

\renewcommand{\theequation}{S.\arabic{equation}}
\setcounter{equation}{0}

\begin{itemize}
	\item \hyperlink{SectionI}{Section I}: A Basis for the $\mathfrak{su}(N)$ Algebra of Qudits
	\item \hyperlink{SectionII}{Section II}: Computation of Structure Constants
	\item \hyperlink{SectionIII}{Section III}: Curvature Calculations
	\item \hyperlink{SectionIV}{Section IV}: The Large $N$ Limit and Sub-Riemannian Geometry
	\item \hyperlink{SectionV}{Section V}: Derivation of the Euler-Arnold Equation for Hydrodynamics
	\item \hyperlink{SectionVI}{Section VI}: Control Hamiltonians, Physical Hamiltonians and Euler-Arnold equations
\end{itemize}
\hypertarget{SectionI}{}\subsection{I. A basis for the $\mathfrak{su}(N)$ algebra of qudits}\label{SectionI}

In this section we explain in detail the construction of a basis for the $\mathfrak{su}(N)$ algebra out of the shift and clock matrices $h$ and $g$, introduced in the beginning of \hyperlink{Section2}{section 2} of the main text. These have the matrix form 
\begin{align}
h &= \begin{pmatrix}
    0 & 1 & 0 & \dots& 0\\
    0 & 0 & 1 & \dots& 0\\
    \vdots & \vdots & \vdots & \ddots& \vdots\\
    0 & 0 & 0 & \dots& 1\\
    1 & 0 & 0 & \dots & 0
    \end{pmatrix},
&
g &= \begin{pmatrix}
    1 & 0 & 0 & \dots& 0\\
    0 & \omega & 0 & \dots& 0\\
    0 & 0 & \omega^2 & \dots& 0\\
    \vdots & \vdots & \vdots & \ddots& \vdots\\
    0 & 0 & 0 & \dots & \omega^{N-1}
    \end{pmatrix},
    \label{shiftnclock}
\end{align}
where we recall that $\omega = {\exp} {2\pi \over N}$. It is readily confirmed that $g$ and $h$ satisfy a modularity condition $g^N=h^N=\mathds{1}$, and commute up to a phase factor, $hg=\omega gh$. The nomenclature for these matrices comes from their action on momentum and position eigenstates in a finite-dimensional Hilbert space. The \enquote{shift} matrix $h$ shifts us from one momentum eigenvector to the next one, while the \enquote{clock} matrix $g$ acts on position eigenstates by multiplying with a phase. This phase is a power of a primitive root of unity, all of which can be visualized as points on the complex unit circle, hence the allusion to a clock. In the continuum limit $N\rightarrow\infty$, these matrices become precisely the position and momentum modes of a plane wave. As introduced in \cite{FAIRLIE1989101,FAIRLIE1989203,Fairlietrigonometric,Patera}, we can use $h$ and $g$ as building blocks to define a basis of unitary, but not necessarily Hermitian, matrices indexed by a two-vector $\vec{m}=(m_1,m_2)$ on the $\mathds{Z}^2$-lattice,
 \begin{equation}
     J_{\vec{m}}=\omega^{\frac{m_1m_2}{2}}g^{m_1}h^{m_2}\,.
     \label{DefinitionoftheJs}
 \end{equation}
We may regard these matrices as non-commutative plane waves, with the vector index $\vec{m}$ playing the role of the wave vector. One can readily verify that the $J$'s are traceless and satisfy $J_{\vec{m}}^{\dagger}=J_{-\vec{m}}$. The $N$-modularity of $h$ and $g$ is inherited by the $J$ matrices and allows the identification of generators on the $\mathds{Z}^2$ lattice via suitable $N$-translations
 \begin{equation}
     J_{\vec{m}+\vec{t}N}=(-1)^{(N t_1 t_2+m_1 t_2+m_2 t_1)}\,J_{\vec{m}}\,,
     \label{Jsymmetry}
 \end{equation}
with $\vec{t}=(t_1,t_2)\in\mathds{Z}^2$ being an integer vector. This divides the $\mathds{Z}^2$-lattice into $N\times N$ cells and allows restriction to generators in what we will denote as the \textit{fundamental cell}, defined by $m_1,m_2=-\frac{N}{2},\cdots,\frac{N}{2}+1$ for even $N$ and $m_1,m_2=-\frac{N-1}{2},\cdots,\frac{N-1}{2}$ for $N$ odd. Note that the odd $N$ case is technically easier to implement due to the symmetry of the fundamental cell around the origin. All numerical computations were thus performed for odd $N$, but the setup is just as valid for even $N$. The generators $J_{(0\,\textrm{mod}\,N,0\,\textrm{mod}\,N)}$ are proportional to the identity and thus decouple from the algebra. The matrices \eqref{DefinitionoftheJs} then close under multiplication to the algebra provided in eq.~\eqref{AlgebraForTheJs} of the main text
 \begin{equation}
     [J_{\Vec{m}},J_{\Vec{n}}]= -2i\sin\left(\frac{\pi}{N}(\Vec{m}\times \Vec{n})\right)\,J_{\Vec{m}+\Vec{n}}\,,
     \label{AlgebraForTheJs'}
 \end{equation}
where $\vec{m}\times\vec{n}\equiv m_1n_2-m_2n_1$.

As they were first introduced in \cite{FAIRLIE1989101,FAIRLIE1989203,Fairlietrigonometric,Patera}, the matrices \eqref{DefinitionoftheJs} are a unitary basis for the $\mathfrak{su}(N)$ algebra thought of as a \textit{vector space}. This construction, however, neglects the relation between Lie algebra elements and elements of the Lie group. In order for a basis of $\mathfrak{su}(N)$ to also act as generators for unitary elements of the Lie group $SU(N)$, it needs to be a set of $N^2-1$ traceless \textit{anti-Hermitian} generators \footnote{We adopt the convention of the mathematical literature and take anti-Hermitian instead of Hermitian matrices.}. To achieve this, we define new generators via the two possible anti-Hermitian linear combinations of $J$'s, one for each vector-index $\vec{m}$,
 \begin{equation}
     C_{\vec{m}}\equiv i(J_{\Vec{m}}+J^{\dagger}_{\Vec{m}})\,,\quad
     S_{\vec{m}}\equiv(J_{\Vec{m}}-J^{\dagger}_{\Vec{m}})\,,
     \label{DefinitionOfTheCsAndSS'}
 \end{equation}
The notation should be suggestive of the usual Euler decomposition for the \textbf{C}osine and \textbf{S}ine functions. However, we have now doubled the number of elements in the basis to $2(N^2-1)$, since every vector in the fundamental cell is associated to both a $C$ and an $S$ matrix. Moreover, some of these $C$'s and $S$'s are linearly dependent.\par

To construct the reduced set of $N^2-1$ linearly independent generators, which we denote as the \textit{final distribution}, we must associate each point in the $(N-1)\times(N-1)$ fundamental cell on the $\mathds{Z}^2$-lattice to an anti-Hermitian generator. In general, we have two possible choices for this generator for each vector $\vec{m}$ in the $\mathds{Z}^2$ lattice, given by $C_{\vec{m}}$ and $S_{\vec{m}}$. Starting from the unique distribution for $N=2$, one can systematically find the distribution for the $(N-1)\times(N-1)$ fundamental cell for $N>2$ by filling the new points of the cell with linearly independent generators, as will be described in the following.\\
Let us first consider the case $N=2$. By construction, $S_{\vec{m}}$ for $\vec{m}\in\{(-\frac{N}{2},-\frac{N}{2}),(-\frac{N}{2},0),(0,-\frac{N}{2}),(-\frac{N}{2},\frac{N}{2}),(\frac{N}{2},\frac{N}{2}),\\(\frac{N}{2},0),(0,\frac{N}{2}),(\frac{N}{2},-\frac{N}{2})\}$ vanish identically for all even $N$. Thus for $N=2$, we have only $C$'s as generators and there is no ambiguity in the distribution within the fundamental cell. For $N=3$, the fundamental cell extends from $-1$ to $1$ in both directions and there are non-vanishing $S$ generators. To insure compatibility \footnote{By \enquote{compatibility} we mean that the structure of the distribution remains the same and subsequent distributions are a natural extension of previous ones. This allows for the systematic generation of the distribution.} with the $N=2$ case, we appoint a $C$ generator to the vectors $(0,1), (1,0)$ and $(1,1)$ of the final distribution. We then fill the remaining points with $S$ generators. However, not all of the remaining points in the $2\times 2$ lattice around the origin host an independent $S_{\vec{m}}$ generator. This is due to the symmetry conditions
\begin{equation}
\begin{split}
     C_{-\vec{m}}&=C_{\vec{m}}\,,\\
     S_{-\vec{m}}&=-S_{\vec{m}}\,.
\end{split}
\label{LinearDependenceConditions}
\end{equation}
We denote \eqref{LinearDependenceConditions} as the \textit{linear dependence condition} (LDC). According to the LDC,  $S_{(1,-1)}\propto S_{(-1,1)}$. Thus, the point $(1,-1)$ must be associated to a $C$ generator, which is guaranteed to be linearly independent from the $S$ by construction.
\begin{figure}
    \centering
   \includegraphics[width=0.45\columnwidth]{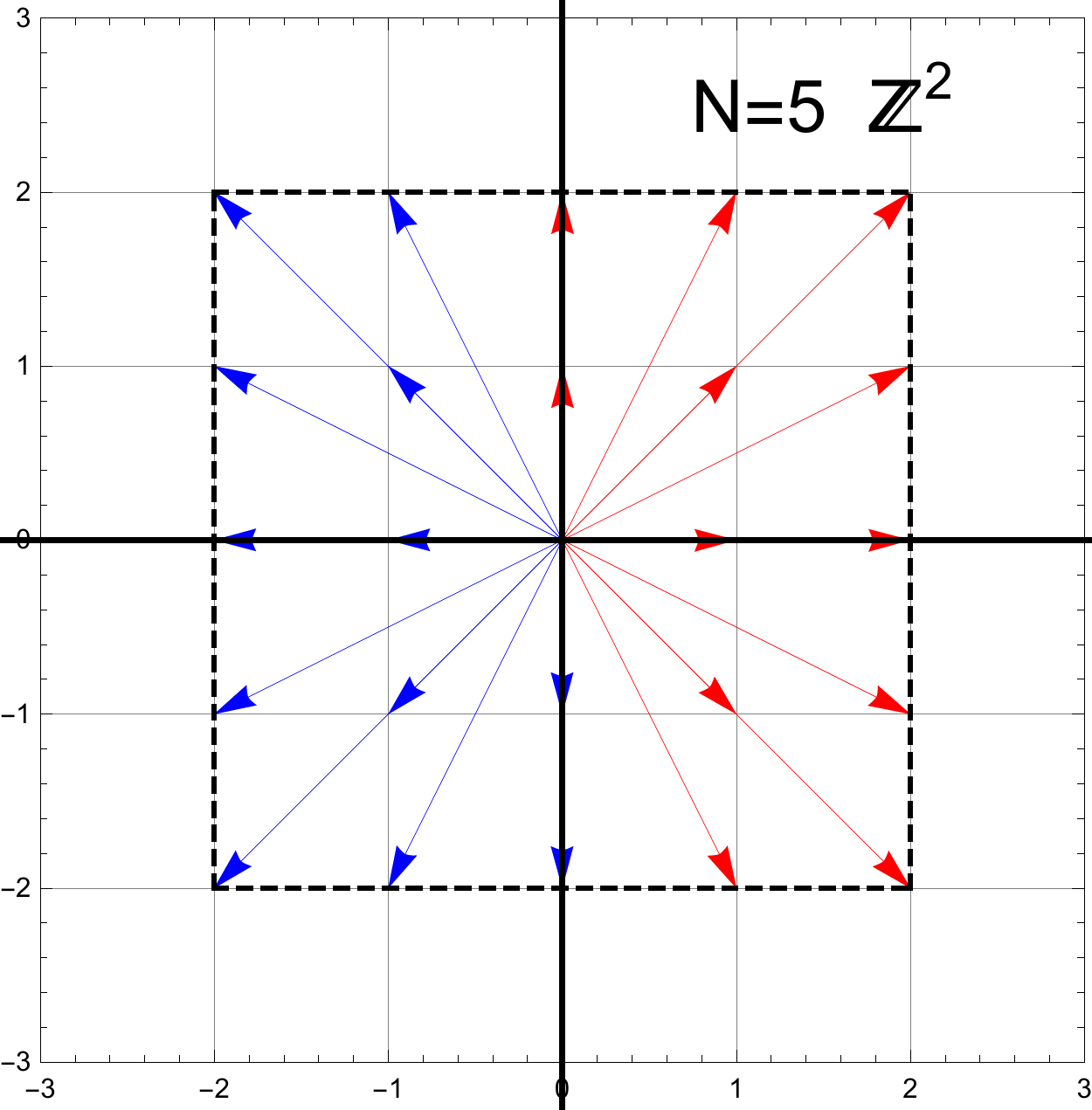}
   \caption{Example for the final distribution of $C$'s (solid red arrows) and $S$'s (dashed blue arrows) inside the fundamental cell (dashed black) for $N=5$.}
   \label{FinalDistributionForN5}
\end{figure}
Once the first two distributions are set, this procedure can be performed iteratively to find the final distribution for $(N+1)$. This construction ensures compatibility with the previous distribution and it completes the remaining points with linearly independent generators according to the LDC symmetry \eqref{LinearDependenceConditions}. In the general odd $N$ case, then, we assign $C$ generators to all points on the right half-plane together with the positive $m_2$-axis and $S$ generators to all points on the left half-plane and the negative $m_2$-axis. A precise list of the points corresponding to $C$'s and those corresponding to $S$'s can be generated automatically for any given $N>3$ (even or odd). As an example, the final distribution for the case $N=5$ is shown in Fig.~\ref{FinalDistributionForN5}.

\hypertarget{SectionII}{}\subsection{II. Computation of Structure Constants}\label{SectionII}
Once the final distribution has been set, as described in the previous section, we can derive the algebra fulfilled by the $C$'s and $S$'s directly from the commutation relation for the $J$'s in \eqref{AlgebraForTheJs'}. These are given by
\begin{align}
    \begin{split}
        [\Tm{C}{m},\Tm{C}{n}]&=2\sin(\frac{\pi}{N}(\vec{m}\times\vec{n}))\Big(C_{\vec{m}+\vec{n}}-C_{\vec{m}-\vec{n}}\Big)\,,\\
        [\Tm{S}{m},\Tm{S}{n}]&=-2\sin(\frac{\pi}{N}(\vec{m}\times\vec{n}))\Big(C_{\vec{m}+\vec{n}}+C_{\vec{m}-\vec{n}}\Big)\,,\\
        [\Tm{C}{m},\Tm{S}{n}]&=2\sin(\frac{\pi}{N}(\vec{m}\times\vec{n}))\Big(S_{\vec{m}+\vec{n}}+S_{\vec{m}-\vec{n}}\Big)\,,\\
        [\Tm{S}{m},\Tm{C}{n}]&=2\sin(\frac{\pi}{N}(\vec{m}\times\vec{n}))\Big(S_{\vec{m}+\vec{n}}-S_{\vec{m}-\vec{n}}\Big)\,.
    \end{split}
    \label{CommutationRelationsForTheCsAndSs}
\end{align}
The mixed commutators in \eqref{CommutationRelationsForTheCsAndSs} seem to not be antisymmetric, due to the relative minus sign inside the bracket on the right-hand side of \eqref{CommutationRelationsForTheCsAndSs}. However, the subtle substitution procedure giving rise to the correct structure constants explained below also ensures that $[\mathcal{T}_{\vec{m}},\mathcal{T}_{\vec{n}}]=-[\mathcal{T}_{\vec{n}},\mathcal{T}_{\vec{m}}]$, with $\mathcal{T}$ either $C$ or $S$, depending on $\vec{m}$.  Therefore, let us now explain how to compute this $(N^2-1)\times(N^2-1)\times(N^2-1)$ anti-symmetric array of structure constants $\ijk{\vec{m}}{\vec{n}}{\vec{l}}$  of the algebra in \eqref{CommutationRelationsForTheCsAndSs}.

There are four possible commutation relations in \eqref{CommutationRelationsForTheCsAndSs}; which one of these is to be used is uniquely determined by the vector indices $\vec{m}$ and $\vec{n}$. Recall that each $\vec{m}$ and $\vec{n}$ is uniquely associated to either a $C$ or an $S$ generator. For example, consider a fixed $\mathds{Z}^2$ vector $\vec{m}$. Then, either $C_{\vec{m}}$ or $S_{\vec{m}}$ will be in the final distribution, but not both. It is practical to define the overall prefactor  
\begin{equation}
    \beta_{(\pm,\pm)}=\pm2\sin(\pm\frac{\pi}{N}(\vec{m}\times\vec{n}))\,,
    \label{betaconstant}
\end{equation}
which is always a part of the structure constants. The subscript $(\pm,\pm)$ denotes which of the four possible sign combinations (overall sign and sign in front of the generator on the RHS of \eqref{CommutationRelationsForTheCsAndSs}, respectively) is to be used. The correct one depends on the commutator and the structure constant we wish to compute. For example, for the $\vec{m}-\vec{n}$ structure constant of the $[C,C]$ commutator, the correct sign combination would be $\beta_{(+,-)}$. Further, the commutator uniquely defines the class of generators on the right-hand side of \eqref{CommutationRelationsForTheCsAndSs}. The correct structure constants are found when the right-hand side is a linear combination of the right class of generators \textbf{and} these are in the final distribution. For example, the right hand side of a $[C,C]$ commutator must be a function of only $C$ generators in the final distribution. However, the $\vec{m}\pm\vec{n}$ indices on the right-hand side of \eqref{CommutationRelationsForTheCsAndSs} might lie outside of the fundamental cell, in which case one has to substitute them on the right-hand side of the commutation relations with their equivalent generators from the final distribution. To relate the generators outside the unit cell to those within it, we use the $N$-modularity of the $J$'s which is inherited by the anti-Hermitian generators as
\begin{equation}
    \Tm{T}{m}=(-1)^{N t_1 t_2+m_1 t_2+m_2 t_1}\Tilde{T}_{\vec{m}+\vec{t}N}\,,
    \label{ModularityConditions}
\end{equation}
for $T\in\{C,S,J\}$. We denote these equations as the \textit{modularity conditions} (MC). Thus, the generators corresponding to integral points on the lattice that can be connected via $N$-translations can also be related. The precise translation required is controlled by the $\vec{t}\in\mathds{Z}^2$ vector and depends on the specific quadrant of the $\mathds{Z}^2$-lattice that one considers. A case-by-case distinction is tedious but can be automatized. Therefore, the overall factor introduced by eq.~\eqref{ModularityConditions} when replacing generators on the right-hand side of the commutation relations is part of the corresponding structure constant $\ijk{\vec{m}}{\vec{n}}{(\vec{m}\pm\vec{n})}$.\\
Note that the point reached after applying the MC condition might not correspond to a generator in the final distribution. In that case, one needs to further apply the LDC condition given in \eqref{LinearDependenceConditions} in order to replace it by the correct generator. This substitution will also carry a prefactor which is to be included into the structure constant.\\
Both MC and LDC conditions are needed to find the final distribution. They are also needed for the computation of the correct structure constants. At first glance it might appear as if \eqref{LinearDependenceConditions} is just a special case of \eqref{ModularityConditions}, but these are indeed independent conditions originating from two distinct symmetries: Eq.~\eqref{ModularityConditions} imposes the modular symmetry of the vectors on the $\mathds{Z}^2$-lattice, while \eqref{LinearDependenceConditions} is a result of having two copies of the original set of generators.\\\par

The overall procedure for calculating the structure constants can be summarized as follows:
\begin{enumerate}
    \item For given input vectors $\vec{m}$ and $\vec{n}$, find the associated generators according to the final distribution and select the corresponding commutator \eqref{CommutationRelationsForTheCsAndSs}. This also defines the class of generators that should appear on the right-hand side of \eqref{CommutationRelationsForTheCsAndSs}. Compute the corresponding generator for $\vec{m}\pm\vec{n}$.
    \item If $\vec{m}\pm\vec{n}$ lies inside the fundamental cell and it corresponds to the right class of generators, the structure constant is simply given by $\beta_{(\pm,\pm)}$ \eqref{betaconstant}.
    \item If $\vec{m}\pm\vec{n}$ lies inside the fundamental cell but (its corresponding generator is) \textbf{not} in the final distribution, apply the LDC condition \eqref{LinearDependenceConditions} and the structure constant is then given by $\beta_{(\pm,\pm)}$ times the prefactor from the LDC condition.
    \item Finally, if $\vec{m}\pm\vec{n}$ does not lie in the fundamental cell, apply the MC condition \eqref{ModularityConditions} to find the corresponding point $\vec{k}$ in the fundamental cell. Repeat steps 2 and 3 with the resulting $\vec{k}$. The structure constant is then given by $\beta_{(\pm,\pm)}$ times the total prefactor from the substitutions.
\end{enumerate}

\hypertarget{SectionIII}{}\subsection{III. Curvature Calculations}\label{SectionVI}
In this section, we elaborate on the computations of sectional and normalized Ricci curvature based on our choice of basis and inner product for the $\mathfrak{su}(N)$. We explain how the sectional curvatures were evaluated in our setup and why a further quantity, the normalized Ricci curvature, is a better quantitative description of geodesic stability.
\subsection{Sectional Curvature}
Sectional curvature $K_{\vec{m}\vec{n}}$ is the scalar curvature of a manifold along the direction of the two-plane spanned by two tangent vectors $\vec{m},\,\vec{n}$  and is defined as \cite{arnold2008topological} 
\begin{equation}
    K_{\vec{m}\vec{n}}=\frac{\inner{R(\vec{m},\vec{n})\vec{m}}{\vec{n}}}{\inner{\vec{m}}{\vec{n}}\inner{\vec{n}}{\vec{n}}}=\frac{1}{m^2n^2}\myroundedbrackets{-\inner{\nabla_{[\vec{m},\vec{n}]}\vec{m}}{\vec{n}}-\inner{\nabla_{\vec{m}}\nabla_{\vec{n}}\vec{m}}{\vec{n}}+\inner{\nabla_{\vec{n}}\nabla_{\vec{m}}\vec{m}}{\vec{n}}}\,.
    \label{WeightedSectCruvFormula}
\end{equation}
Using our choice of inner product given in eq.~\eqref{InnerProduct} of the main text and the commutation relations \eqref{CommutationRelationsForTheCsAndSs} resulting from our choice of basis, we can derive the explicit form of the sectional curvatures tensor in our setup to be
\begin{equation}
    \begin{split}
        K_{\vec{m}\vec{n}}=\frac{1}{m^2n^2}\Bigg(\sum_{\vec{k}}&\frac{1}{2}\ijkvec{m}{n}{l}\myroundedbrackets{-\ijkvec{m}{n}{l}\,l^2+\ijkvec{n}{l}{m}m^2+\ijkvec{l}{m}{n}n^2}\\
        &-\frac{1}{4k^2}\myroundedbrackets{\ijkvec{m}{n}{k}k^2-\ijkvec{n}{k}{m}m^2+\ijkvec{k}{m}{n}n^2}\myroundedbrackets{\ijkvec{m}{n}{k}k^2+\ijkvec{n}{k}{m}m^2-\ijkvec{k}{m}{n}n^2}\\
        &-\frac{1}{k^2}\ijkvec{k}{m}{m}\ijkvec{k}{n}{n}m^2n^2\Bigg)\,.
    \end{split}
    \label{SectionalCurvatureTensor}
\end{equation}
\begin{figure}[t]
    \centering
    \includegraphics[width=0.75\columnwidth]{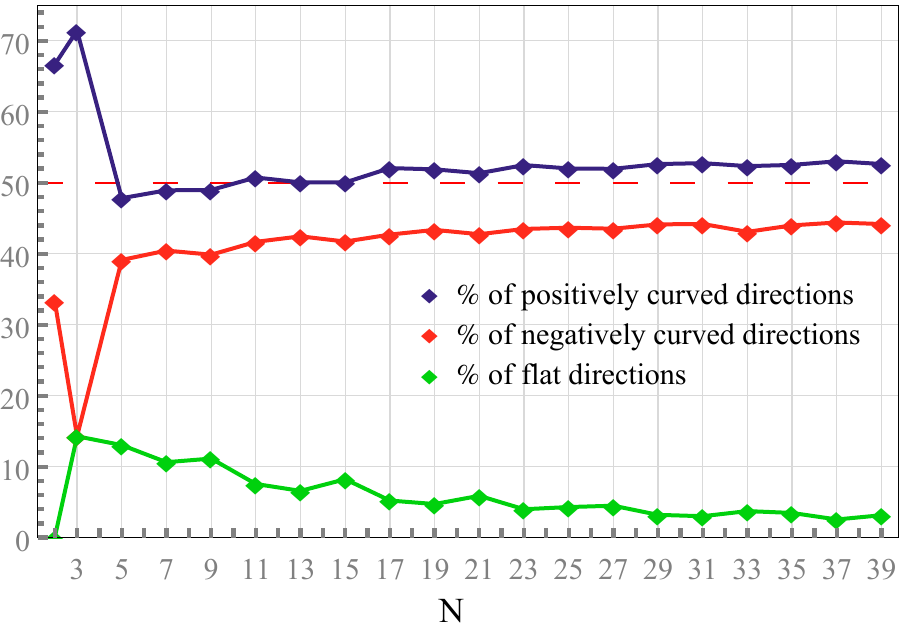}
    \caption{Distribution of signs of sectional curvatures of $SU(N)$ in directions given by pairs of generators of the $\mathfrak{su}(N)$ algebra for $2\leq N\leq 39$. The computations exhibited changes of less than $0.1\%$ for large $N$, strongly suggesting that the percentages stabilize. Lines are a visual guide, not an interpolation. A slight majority of positively curved directions settles, but this is inconclusive with regards to the stability of geodesics on the manifold of unitaries.}
    \label{PlotSectCurves}
\end{figure}
Based on our finite-dimensional setup, we evaluated \eqref{SectionalCurvatureTensor} for odd values of $N\in[2,39]$. The value $N=2$ is simple enough to be carried out analytically and was thus taken into account. The resulting percentages of non-trivially flat~\footnote{Trivially flat directions are those given by commuting generators.}, negatively and positively curved directions between the generators \eqref{DefinitionOfTheCsAndSS'} are shown in Fig.~\ref{PlotSectCurves}. We find that the percentages seemingly stabilize for large $N$ to values of $4\,\%$, $44\,\%$, and $52\,\%$ respectively. Although it seems as if the flat directions will vanish, this is just an artifact. There will always be non-trivial vanishing structure constants, and thus flat sectional curvatures, since for all $N>2$ there will be at least $N-1$ vectors in the unit cell which are linearly dependent (but whose corresponding generators are not) that yield identically vanishing structure constants due to $\vec{m}\times\vec{n}$. Note that the result for the single-qubit case $N=2$, where there are only three non-trivial directions, we find a ratio of $\frac{1}{3}:\frac{2}{3}$ of negative to positive sectional curvatures, which coincides with previous analysis of complexity for other cost functions \cite{Auzzi:2020idm}.

The interest of previous works in sectional curvatures of complexity metrics lies in their role as indicator for geodesic instability.  In particular, if the majority of sectional curvatures in the tangent space is negative, nearby geodesics have a high probability of exponentially diverging and thus lead to classically chaotic trajectories. Unfortunately, our numerical data on the distribution of the sectional curvatures on $SU(N)$ with respect to our complexity metric, though robust, is inconclusive as to whether typical trajectories on the manifold of unitaries will behave chaotically. This is not in contradiction with the results of Arnold \cite{arnold2008topological}, who showed that \eqref{EAEquation} evaluated on SDiff($\mathds{T}^2$) hosts a large class of solutions with negative sectional curvature. To see this, we note that while Arnold concluded SDiff($\mathds{T}^2$) is, in his words, \enquote{mostly negative} \cite{arnold2008topological}, he does not provide a quantitative description of the sectional curvatures or a rigorous, measure-theoretic description of his results.

 
\subsubsection{Normalized Ricci curvature}

The present subsection is concerned with evaluating the normalized or average Ricci curvature introduced in \hyperlink{Section5}{section 5} of the main text, which we find to be a better measure of stability than the sectional curvature alone.

Using our results for the sectional curvatures, we use Eq.~\eqref{NormalizedRicciCurvature} of the main text to compute the normalized Ricci curvature for every direction in the $\mathfrak{su}(N)$ algebra and track its evolution with $N$ up to $N=39$. Through this calculation, we are able to calculate the critical $N_c$ of each direction, at which the normalized Ricci curvature becomes negative. The results are shown in Fig.~\ref{MainPlot} of the main text. 

This figure can be used to divide the generators into $\oforder{1}$ and $\oforder{N}$ subsectors. We observe from Fig.~\ref{MainPlot} of the main text, that the value of $N_c$ seems to grow very slowly with the distance of the vector to the origin. However, it remains finite. This means that every possible direction on the $\mathfrak{su}(N)$ algebra for finite $N$ will initially have a positive normalized Ricci curvature, but will eventually turn negative at a higher, finite value of  $N$. For any finite $N$, there will always be a next higher $N$ at which new directions will enter the final distribution and will have a positive Ricci curvature. These belong to the $\oforder{N}$ subsector. Simultaneously, there will be directions whose curvature will have already reached the critical value and have turned negative. These will belong to the $\oforder{1}$ sector. However, in the strict large $N$ limit, a transition to an infinite-dimensional manifold takes place. This $SU(\infty)$ manifold will have an $\oforder{1}$ subsector, which contains all the directions that can be identified isomorphically with Hamiltonian vector fields. Indeed, this is the entirety of the hydrodynamic phase-space, corresponding to all vectors on $\mathds{Z}^2$.

\hypertarget{SectionIV}{}\subsection{IV. The Large $N$ Limit and Sub-Riemannian geometry}\label{SectionIII}

As mentioned in the main text, the isomorphism identifying generators of $\mathfrak{su}(N)$ and Hamiltonian vector fields at large $N$ is subtle. Some families of vectors have cross products of order $\oforder{N}$ or $\oforder{N^2}$ large enough to impede a truncation of the sine to first order. In these cases, an identification with Hamiltonian vector fields is not proper. However, for all vectors whose cross product is of order $\oforder{1}$, the argument of the sine is indeed small and a truncation of the Taylor expansion can be made. This restricts the implementation of the isomorphism to only those vectors in this $\oforder{1}$ subsector. 

In terms of qudit unitary gates, our basis allows the implementation of any gate describing the transition between two energy levels inside the qudit, cf. fig.~\ref{QuditLevels}. However, the restriction to low-momentum $\oforder{1}$ vectors means that the large $N$ limit, and thus, the hydrodynamic description, impose a certain notion of locality. Though it is not a hard cut-off locality prescription, such as the notion of $k$-locality often imposed when considering complexity of many-qubit gates in the Pauli basis, this constraint indicates that gates implementing $\oforder{N}$ are highly penalized while $\oforder{1}$ gates are preferred in terms of cost. Of course, the precise meaning of $\oforder{1}$ and $\oforder{N}$ changes for every finite value of $N$. Some vectors which are in the $\oforder{N}$ sector for a given $N$ will be contained in the $\oforder{1}$ sector for a higher $N$, which gives rise to the evolution of Ricci curvatures explained in section \hyperlink{SectionIII}{III} of the supplemental and section \hyperlink{Section5}{5} of the main text.

The situation explained above can be described within the framework of \textit{sub-Riemannian geometry} \cite{agrachev2019}. The relevance of sub-Riemannian geometry for complexity was originally mentioned by Nielsen \cite{NielsenLowerBounds} and has been recently invoked again by Susskind and collaborators \cite{Brown:2021rmz}. Our setup can be thought of as defining a family of metrics $\{(g_{\mu\nu})_N\}$ for every finite $N$. Each of these metrics will impose a weak locality condition by applying larger penalties to infinitesimal gates in $\oforder{N}$ directions. The transition to the strict large $N$ limit carries a transition to infinite-dimensional manifold with it. In this limit, the family of smooth, approximately sub-Riemannian metrics defined above converge to a proper sub-Riemannian metric on the $SU(\infty)$ manifold,
\begin{equation}
	\lim_{N\rightarrow\infty}\{(g_{\mu\nu})_N\}=\mathfrak{G}_{\mu\nu}\,.
\end{equation}
In particular, this indicates that hydrodynamics emerges in the strict large $N$ limit as the low-energy, $\mathcal{O}(1)$ \textit{sub-Riemannian manifold} of the much larger $SU(N\rightarrow\infty)$ group manifold. The metric $\mathfrak{G}_{\mu\nu}$ on the latter consists of two objects: First, the Laplacian, which applies a penalty $m^2$ in all directions $\vec{m}$ corresponding to the hydrodynamical phase-space. Note that this encompasses all \textit{finite} penalties available on $\mathds{Z}^2$. Second, a \textit{hard-wall} metric, imposing infinite penalties in all remaining directions which arose from the strict large $N$ limit of the $\oforder{N}$ sector. These directions are denoted as \textit{inadmissible}.

The notion of \textit{controllability} is fundamental to Nielsen's geometric approach to complexity. It implies that every point of the manifold can be reached via a geodesic following the admissible directions defined by the cost function. In this regard, a main result of sub-Riemannian geometry is the Chow-Rashevskii theorem \cite{Rashevskii1938,Chow1940}, which guarantees controllability on finite-dimensional manifolds provided that the basis for the tangent space is a \textit{bracket-generating distribution}. This means that every direction in the tangent bundle can be equivalently described by a finite chain of commutators involving only generators associated to admissible directions. For any finite $N$, our basis \eqref{DefinitionOfTheCsAndSS'} fulfills the bracket-generating condition, even when restricting to the $\oforder{1}$ subsector. This is because $\mathfrak{su}(N)$ is a simple Lie algebra. However, the proper sub-Riemannian metric $\mathfrak{G}_{\mu\nu}$ is defined only in the strict large $N$ limit. Fortunately, an analogue of the Chow-Rashevskii theorem on infinite-dimensional manifolds exists \cite{KhajehSalehani2014ControllabilityOI}. This suggests that controllablity survives the transition to infinite $N$ in our construction. In other words, we can reach any point of the large $SU(\infty)$ manifold via geodesics that flow only through admissible directions given by the hydrodynamic phase-space. Thus, the computation of Nielsen complexity as explained in the main text and in section \hyperlink{SectionVI}{VI} proves to be well-posed.

 The growth of the penalty factors with $N$, implies that it becomes increasingly more ``expensive" to move directly in an $\oforder{N}$ direction, as we veer further and further away from the identity operator, in terms of geodesic distance on $SU(N)$. Instead, the system will prefer to move indirectly towards an $\oforder{N}$ direction through the ``cheaper" $\oforder{1}$ directions \footnote{The Chow-Rashevskii theorem ensures this is possible.}. This heuristic understanding is made more quantitative via the recent conjecture of \cite{Brown:2021rmz}. This conjecture involves a family of metrics, of the same dimension, $\{(g_{\mu\nu})_{\cal I}\}$, indexed by the penalty factor of the hard directions, ${\cal I}$ \footnote{Here we have assumed all the hard directions are degenerate for simplicity. The more general case is treated in \cite{Brown:2021rmz}.}. Namely, the conjecture states that for large values of ${\cal I}$ and long geodesic distances 

\begin{equation}
\label{Eq:Conjecture} {\cal C}_{{\cal I} =\infty} = {\cal C}_{{\cal I}} + {\cal O}\left({1\over \sqrt{{\cal I}}}\right)~.
\end{equation}
The index ${\cal I}$ can also be implicitly defined from the leading order behaviour of the sectional curvatures for large ${\cal I}$, schematically $K \sim {\cal I}$. In our case, we can confirm both through the metric and our numerical calculation of the sectional curvatures that, ${\cal I}$ is identified with $N^2$. With this identification, Eq. \eqref{Eq:Conjecture} leads to the relation between ${\cal C}(SU(N))$ and ${\cal C}$(SDiff($\mathds{T}^2$)) mentioned at the end of section \hyperlink{Section5}{5} in the main text. 

Note that, we use the conjecture of Susskind et al. \cite{Brown:2021rmz} for a family of metrics of different dimension. This is not a violation of the conjecture, but rather a confirmation of the expectations of the authors as seen in the discussion section of \cite{Brown:2021rmz}.

A further aspect of this conjecture is the existence of a critical ${\cal I}$, for which the long geodesic distance behaviour matches the short one. At this critical value, it is in general easier to calculate complexity and confirm explicitly the leading scaling behaviour \eqref{Eq:Conjecture}. On that note, let us expand on the comment made in the \hyperlink{Section6}{conclusions section} of the main text regarding an explicit derivation of \eqref{Eq:Conjecture}. A rigorous derivation of such scalings requires a more detailed analysis of the isomorphism relating $\mathfrak{su}(N\rightarrow\infty)$ and SVect($\mathds{T}^2$). This is achieved by considering a non-commutative version of plane waves. The mathematical framework formalizing this problem is that of non-commutative geometry (NCG) \citep{connes1995noncommutative}.  The standard procedure is to introduce the so-called Weyl-Moyal star product \cite{Khesin_2004},
\begin{equation}
	(f\star g)(x,p):=fg+\sum_{n=1}^{\infty} \frac{(\pi \theta)^n}{n!}\epsilon_{r_1s_1}\dots \epsilon_{r_ns_n}(\partial^n_{r_1\dots r_n}f)(\partial^n_{s_1\dots s_n}g)\,.
\label{WeylMoyalProduct}
\end{equation}
This defines a way of multiplying classical smooth functions in a non-commutative manner. The deformation parameter is $\theta=1/N$ in our case. This leads to a deformed Poisson bracket, the Moyal bracket $\{f,g\}_{\star}$, which in turn defines non-commutative algebras for vector fields. This also yields a non-commutative version of the Euler-Arnold equation
\begin{equation}
	\Delta \dot{f}=-\{f,\Delta f\}_{\star}\,.
\end{equation}
In order to properly investigate the scaling of Nielsen complexity on $SU(N)$ in terms of that on SDiff($\mathds{T}^2$), we have to determine higher orders in the algebra isomorphism given in eq.~\eqref{AlgebraIsomoprhism} of the main text. These are encoded in the expansion of the Weyl-Moyal product, which provides the higher-order in $1/N$ corrections to the commutative Euler-Arnold equation given in eq.~\eqref{EAEquationHydro} of the main text. A proper identification of finite- and infinite-dimensional algebras at higher orders in $1/N$ would provide a scaling relation of the type discussed here. There have also been attempts of discretizing the Laplacian in non-canonical ways as a way to define finite-dimensional approximations to two-dimensional hydrodynamics \cite{PhysRevA.46.6417}, which would imply the definition of a new cost function in Nielsen's approach. We leave such investigations for future work.

\hypertarget{SectionV}{}\subsection{V. Derivation of the Euler-Arnold equation for hydrodynamics}\label{SectionIV}
The derivation of the Euler-Arnold equation for the hydrodynamics system  given by eq.~\eqref{EAEquationHydro} in the main text relies on the following steps. We use Arnold's conventions \cite{arnold2008topological} and define Hamiltonian vector fields as $\iota_{X_f}\omega = -\mathbf{d}f$, with $\omega$ the symplectic form on $\mathds{T}^2$ and $f$ the Hamiltonian function dual to $X_f$. $\mathbf{d}$ is the exterior derivative and $\iota$ denotes the interior product. The symplectic form further defines the Poisson bracket of Hamiltonian functions as $\{f,g\}=\omega(X_f,X_g)$. This implies that the commutator of Hamiltonian vector fields is itself a Hamiltonian vector field. This new vector field has Hamiltonian function given by the Poisson bracket of the Hamiltonian functions $f$ and $g$,
\begin{equation}
    \begin{split}
    \iota_{[X_f,X_g]}\omega&=\mathscr{L}_{X_f} \iota_{X_g} \omega-\iota_{X_g}\mathscr{L}_{X_f} \omega\\
    &=\mathbf{d}\iota_{X_f}\iota_{X_g}\omega+\iota_{X_f}\underbrace{\mathbf{d}\iota_{X_g}\omega}_{=0}-\iota_{X_g} \underbrace{\mathbf{d}\iota_{X_f}\omega}_{=0}-\iota_{X_g}\iota_{X_f}\underbrace{\mathbf{d}\omega}_{=0}\\
    &=-\mathbf{d}\omega(X_f,X_g)=-\mathbf{d}(\{f,g\}) = \iota_{X_{\{f,g\}}}\omega \,.
    \end{split}
    \label{HamiltonianFunctionOfTwoSymplVectorFields}
\end{equation}
Here, $\mathscr{L}_{Y}h=Y(h)$ denotes the Lie derivative of a smooth function $h$ in the direction of the vector field $Y$ and we made use of Cartan's magic formula $\mathscr{L}_{X}=\mathbf{d}\,\iota_X+ \iota_X \mathbf{d}$. We now use relation \eqref{HamiltonianFunctionOfTwoSymplVectorFields}, together with the definition of the $L^2$-inner product in eq.~\eqref{L2Product} of the main text, to compute the $\kappa$ form
\begin{equation}
    \begin{split}
         \left<[X_f,X_g],X_f\right>_{\mathbf{T}^2} &= \left<X_{\{f,g\}},X_f\right>_{\mathbf{T}^2}\\
         &= -\int_{\mathbf{T}^2} \{f,g\}\Delta f\\
         &= \int_{\mathbf{T}^2} g \Delta \Delta^{-1}\{f,\Delta f\}\\
         &\overset{!}{=}\left<X_g,\kappa(f,f)\right>_{\mathbf{T}^2}\,.
    \end{split}
    \label{DerivationKappaEulerArnold}
\end{equation}
The first equality in \eqref{DerivationKappaEulerArnold} implements \eqref{HamiltonianFunctionOfTwoSymplVectorFields}. The insertion of the definition of the  $L^2$-inner product then yields the second line. To arrive to the third line, we make use of the Leibniz rule for the Poisson bracket and the fact that $\{f,g\}\omega$ can be rewritten as the exterior derivative of a 1-form $\mathbf{d}(g\,\iota_{X_f}\omega)$ via Cartan's formula, for any Hamiltonian functions $f$ and $g$. Then, the generalized Stokes' theorem says that $\int_{\mathcal{M}}\mathbf{d}\alpha=\int_{\partial \mathcal{M}}\alpha$ for any 1-form $\alpha$. For the torus,  $\partial \mathcal{M}=0$ and, hence the integral over $\alpha$ vanishes identically. In the second integral of the third line of \eqref{DerivationKappaEulerArnold}, we insert an identity operator in the form of $\Delta\Delta^{-1}$, to recover the expression for the $L^2$-inner product. From this we can read off the resulting $\kappa$ form
\begin{equation}
\kappa(f,f)=-\Delta^{-1}\{f,\Delta f\}\,,
\end{equation}
which directly determines the EA equation given by eq.~\eqref{EAEquationHydro} in the main text.

\hypertarget{SectionVI}{}\subsection{VI. Control Hamiltonians, Physical Hamiltonians and Euler-Arnold equations}\label{SectionV}
We now detail the precise role of the control Hamiltonian in the computation of complexity and the technical difficulties that arise in its calculation. First, we clarify some notation: even though Nielsen's approach is valid for general unitaries, a usual choice for the target unitary $U$ in the context of holography is the time evolution operator $U=e^{-iHt}$ of the system under some \textit{physical} Hamiltonian $H$ up to time $t$. Control Hamiltonians $\mathcal{H}(s)$ and physical Hamiltonians $H$ are distinct objects; the former generates a trajectory, while the latter specifies the boundary condition to this trajectory, meaning that we impose $U(1)=e^{-iHt}$ on the geodesics generated by some $\mathcal{H}$. We note that the geodesic parameter $s$ is not necessarily identical with the physical time $t$. The computation of complexity can be generally divided into two main steps: the computation of possible control Hamiltonians and the implementation of boundary conditions to find those that actually generate the desired unitary, i.e. solving the Schrödinger equation. We now review these steps. 

Within the Euler-Arnold approach \cite{AIF_1966__16_1_319_0,arnold2008topological} introduced in the main text, the defining equation for the control Hamiltonian is the so-called Euler-Arnold equation \cite{NielsenQCasGeometry},
\begin{equation}
	\dot{\mathcal{H}}(s)=\kappa(\mathcal{H}(s),\mathcal{H}(s))\,,
	\label{EAEquation}
\end{equation}
with the quadratic bilinear form $\kappa$ defined on the Lie algebra as
\begin{equation}
	\langle[X,Y],Z\rangle=\langle\kappa(Z,X),Y\rangle\,,
\end{equation}
and $\langle\cdot,\cdot\rangle$ the inner product on the algebra. 
Solutions to \eqref{EAEquation} in principle yield direct expressions for the control Hamiltonian. However, such a direct computation of the Hamiltonian is in general rather involved, since the $\kappa$ form with respect to metrics other than the canonical Killing metric yield complicated differential equations. Moreover, the subsequent evaluation of the Schrödinger equation contains expressions with nested commutators \cite{Magan:2018nmu}, which are generally difficult to evaluate. For this reason, the usual procedure is to change perspective and expand the control Hamiltonian in a basis, in order to obtain more tractable equations for the expansion coefficients. On the manifold of unitaries, a chart is provided by a basis of the tangent space,  given by anti-Hermitian generators. Due to the group structure of the manifold, every tangent space is isomorphic to the Lie algebra, and thus we can expand the control Hamiltonian in terms of our generators of $\mathfrak{su}(N)$,
\begin{equation}
	\mathcal{H}(s)=\sum_{\vec{m}\in\mathfrak{su}(N)}^{N^2-1}Y^{\vec{m}}			(s)\mathcal{T}_{\vec{m}}\,
	\label{HamiltonianExpansion},
\end{equation}
for $\mathcal{T}\in\{C,S\}$ according to the final distribution.
Note that the expansion coefficients $Y^{\vec{m}}(s)$  encode the time-dependence of the Hamiltonian, while the generators are fixed. The coefficients can be thought of as velocities and are formally given by
\begin{equation}
	Y^{\vec{m}}(s)\mathcal{T}_{\vec{m}}=\frac{d U(s)}{ds}U(s)^{-1}\,.
\end{equation}
These velocities obey a geodesic equation with respect to the metric associated to the cost function on $SU(N)$, i.e. our inner product defined in eq.~\eqref{InnerProduct} of the main text. Solutions will yield the initial velocities as a set of integration constants. These have to be inserted into the Schrödinger equation and the boundary conditions need to be imposed. This way we can relate the initial velocities to the actual target unitary, i.e. we can find $Y^{\vec{m}}=Y^{\vec{m}}(s,U)$. In the case of unitary time-evolution, this step will relate the velocities to the couplings in the physical Hamiltonian of the system, which can also be expanded in terms of the generators of the $SU(N)$ group as $H=\sum_{\vec{m}}A^{\vec{m}}\mathcal{T}_{\vec{m}}$. In the end, one wishes to have a set of functions $Y^{\vec{m}}(s,A^{\vec{k}})$ which define the correct initial velocities of the geodesic. In summary, the overall procedure for computing quantum complexity á la Nielsen consists of the following steps.
\begin{enumerate}
	\item Choose a cost function $\mathcal{F}$ which will define a metric on the manifold of unitaries, thus also defining geodesic paths with respect to it. In our case this is given by the inner product on $\mathfrak{su}(N)$ given by eq.~\eqref{InnerProduct} of the main text.
	\item Derive the geodesic equation for this metric. This can be in terms of the control Hamiltonian $\mathcal{H}$ via the Euler-Arnold equation \eqref{EAEquation} or in terms of the geodesic equation for the velocities $Y$ that determine $\mathcal{H}$. Which approach is to be taken depends on the system in consideration. Solutions to these equations yield a family of control Hamiltonians with integration constants that need to be fixed by the boundary conditions to the trajectory.
	\item Insert these control Hamiltonians into the Schrödinger equation $\frac{dU}{ds}=\mathcal{H}U$ and enforce the correct boundary conditions, i.e. $U(0)=\mathds{1}$ and $U(1)=U_{\textrm{target}}$. This will fix the velocities of the geodesic, through which the length can be calculated via the standard length functional of Riemannian geometry with respect to the chosen metric.
	\item If multiple control Hamiltonians are found, one has to choose the one which minimizes the length, and thus the cost. This yields the complexity $\mathcal{C}$.
\end{enumerate}

The benefit arising from our analysis is that, in the large $N$ limit, we identify the $\mathfrak{su}(N)$ algebra with the algebra of volume-preserving diffeomorphisms SVect($\mathds{T}^2$) via the isomorphism between generators in eq. \eqref{AlgebraIsomoprhism} of the main text. This naturally carries over to any other tangent space, such that we can, locally and at low energies, identify the $SU(N)$ manifold with the manifold of volume-preserving diffeomorphisms SDiff($\mathds{T}^2$). Our results show how to define the general setup for Nielsen complexity explained above in such a way that it survives the large $N$ limit. The whole construction can be interpreted in terms of objects in two-dimensional hydrodynamics. Most importantly, our setup simplifies the first step of the computation considerably, providing a tractable and well-known Euler-Arnold equation for the control Hamiltonian, namely the Euler equation for ideal hydrodynamics in (2+1) dimensions given in eq.~\eqref{EAEquationHydro} of the main text. The control stream function $\mathpzc{h}$ and the corresponding control Hamiltonian vector field $\mathcal{H}$ are related in the usual way, $\iota_{\mathcal{H}}\omega=-\mathbf{d}\mathpzc{h}$. It is important to note that both the control stream function and vector field have an implicit dependence $(x,p)$ on the coordinates of the fluid on the torus. Solutions of this equation can then uniquely be decomposed in order to find the velocities with which the length functional can be evaluated. Naturally, the correct boundary conditions need to be imposed. The Schrödinger equation in the hydrodynamic setup corresponds to the mapping between Lagrangian and Eulerian frames of reference for the fluid \cite{goldstein2002classical},
\begin{equation}
	\frac{df(s)}{ds}=\mathcal{H}(s,f(s))\,,
\end{equation}
and relates the control Hamiltonian vector field $\mathcal{H}(s)$ with the diffeomorphism $f$ it generates. The appropriate boundary conditions should then be imposed on this equation, in particular $f(1)=f_{\textrm{target}}$, where the target diffeomorphism may be taken to be the time evolution of the fluid under some physical Hamiltonian, which we describe in the next section.

\subsubsection{Physical Hamiltonian}
We now discuss some details on the role of physical Hamiltonians in Nielsen complexity. Even though the approach explained above and our results are valid for any target operation $U$, the literature on holographic complexity generally focuses on the target unitary being the time evolution operator. In recent years it has been found \cite{susskind2014entanglementisnotenough} that entanglement is not enough to describe all the late time properties of a holographic QFT and the inner black hole geometry dual to it. Thus, complexity was proposed \cite{Brown:2016wib} as a suitable additional measure of the information content of the QFT at late times. To focus on the late time behavior, one considers the unitary operation to be the time evolution operator $U=e^{-iHt}$ of the system at time $t$ under some physical Hamiltonian $H$ and calculates its complexity. This physical Hamiltonian is usually assumed to be $k$-local or chaotic. Thus, we detail some properties that such a Hamiltonian should have if it ought to be compatible with our setup.
\begin{figure}[h!]
	\includegraphics[scale=0.35]{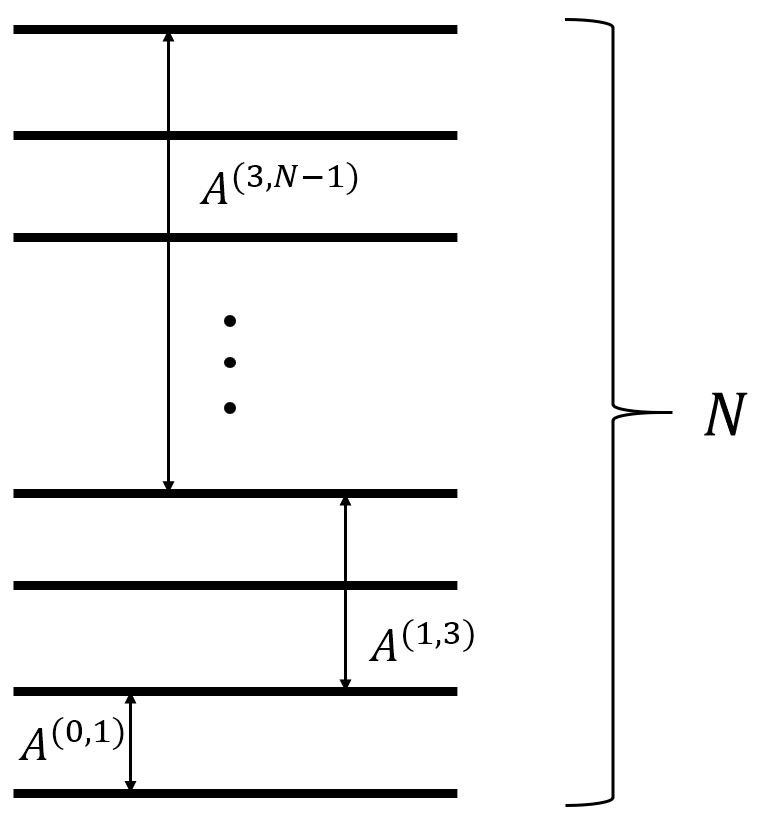}
	\caption{Internal qudit levels with exemplary transitions. The physical Hamiltonian $H$ is assumed to implement all-to-all transitions, each of which is determined, up to eight-fold degeneracy, by the wave vector of the associated excitation.}
	\label{QuditLevels}
\end{figure}

The physical system we consider is given by a single $N$-level qudit, see Fig.~\ref{QuditLevels}. Notions of locality such as $k$-locality may be interpreted in this system as the type of internal transitions between the levels that are allowed by the Hamiltonian, e.g. only transitions with energy smaller than some threshold $k$. However, we take a different approach. We allow the physical Hamiltonian to contain, at first, all possible couplings $A^{\vec{m}}$ between levels, i.e. $H_{\textrm{qudit}}$ implements all-to-all transitions on the qudit. Written in terms of generators of the algebra in the final distribution,  we have 
\begin{equation}
	H_{\textrm{qudit}}= \sum_{\vec{m}}^{N^2-1}A^{\vec{m}}\mathcal{T}_{\vec{m}}\,,\quad \quad \mathcal{T}\in\{C,S\}\,.
\end{equation}
A natural assumption would be that there exists no hierarchy between the couplings, i.e. all couplings are assumed to be of order $\mathcal{O}(1)$. It is worth noting that although we do not impose any locality restrictions on the Hamiltonian itself, our choice of cost function has an approximate locality property built into the penalties, which is responsible for the restriction to the $\mathcal{O}(1)$ sector.

Similarly to the previous discussion for the qudit Hamiltonian, we can interpret these assumptions for volume-preserving diffeomorphisms. More precisely, we can choose the target diffeomorphism $f_{\textrm{target}}$ introduced in the previous section to be the time evolution of the fluid under some physical Hamiltonian $H_{\textrm{fluid}}$ after some time $t$. We assume this Hamiltonian to obey similar properties as $H_{\textrm{qudit}}$, albeit adapted to the infinite-dimensional Hilbert space. In particular, this means there are infinite many levels, each one excited by a particular plane wave with wave vector $\vec{m}$. The components of this wave vector remain integer, as they have to be compatible with the periodicity of the torus. This Hamiltonian is still assumed to implement all-to-all interactions.  Expanding again in terms of the generators, this time of SDiff($\mathds{T}^2$), we have the schematic structure
\begin{equation}
	H_{\textrm{fluid}}=\sum_{f} B^{f}\cdot X_f\,,
\end{equation}
with the couplings $B^f$ indexed by the corresponding plane wave $f_{\vec{m}}$ with wave vector $\vec{m}$. For our analysis, we do not assume a direct connection between $A^{\vec{m}}$ and $B^{f}$, instead leaving them free, subject only to the general assumptions explained above. 
\end{widetext}

\bibliographystyle{ieeetr}
\bibliography{Refs.bib}

\begin{thebibliography}{10}

\bibitem{watrous2008quantum}
J.~Watrous, ``Quantum computational complexity,'' 2008.

\bibitem{NielsenLowerBounds}
M.~A. Nielsen, ``A geometric approach to quantum circuit lower bounds,'' {\em
  Quantum Info. Comput.}, vol.~6, p.~213–262, May 2006.

\bibitem{NielsenQCasGeometry}
M.~A. Nielsen, M.~R. Dowling, M.~Gu, and A.~C. Doherty, ``Quantum computation
  as geometry,'' {\em Science}, vol.~311, no.~5764, pp.~1133--1135, 2006.

\bibitem{NielsenGeometryOfQC}
M.~R. Dowling and M.~A. Nielsen, ``The geometry of quantum computation,'' {\em
  Quantum Info. Comput.}, vol.~8, p.~861–899, Nov. 2008.

\bibitem{PhysRevLett.122.231302}
P.~Caputa and J.~M. Magan, ``{Quantum Computation as Gravity},'' {\em Phys.
  Rev. Lett.}, vol.~122, no.~23, p.~231302, 2019.

\bibitem{erdmenger2020complexity}
J.~Erdmenger, M.~Gerbershagen, and A.-L. Weigel, ``{Complexity measures from
  geometric actions on Virasoro and Kac-Moody orbits},'' {\em JHEP}, vol.~11,
  p.~3, 2020.

\bibitem{Flory:2020eot}
M.~Flory and M.~P. Heller, ``{Geometry of Complexity in Conformal Field
  Theory},'' {\em Phys. Rev. Res.}, vol.~2, no.~4, p.~043438, 2020.

\bibitem{Flory:2020dja}
M.~Flory and M.~P. Heller, ``{Conformal field theory complexity from
  Euler-Arnold equations},'' {\em JHEP}, vol.~12, p.~091, 2020.

\bibitem{Maldacena:1997re}
J.~M. Maldacena, ``{The Large N limit of superconformal field theories and
  supergravity},'' {\em Adv. Theor. Math. Phys.}, vol.~2, pp.~231--252, 1998.

\bibitem{Witten:1998qj}
E.~Witten, ``{Anti-de Sitter space and holography},'' {\em Adv. Theor. Math.
  Phys.}, vol.~2, pp.~253--291, 1998.

\bibitem{Gubser:1998bc}
S.~S. Gubser, I.~R. Klebanov, and A.~M. Polyakov, ``{Gauge theory correlators
  from noncritical string theory},'' {\em Phys. Lett. B}, vol.~428,
  pp.~105--114, 1998.

\bibitem{Note1}
See \cite {Haferkamp:2021uxo} for a recent proof of this linear growth of
  complexity for Haar-random circuits.

\bibitem{Brown:2016wib}
A.~R. Brown, L.~Susskind, and Y.~Zhao, ``{Quantum Complexity and Negative
  Curvature},'' {\em Phys. Rev. D}, vol.~95, no.~4, p.~045010, 2017.

\bibitem{Brown:2017jil}
A.~R. Brown and L.~Susskind, ``{Second law of quantum complexity},'' {\em Phys.
  Rev. D}, vol.~97, no.~8, p.~086015, 2018.

\bibitem{Brown:2019whu}
A.~R. Brown and L.~Susskind, ``{Complexity geometry of a single qubit},'' {\em
  Phys. Rev. D}, vol.~100, no.~4, p.~046020, 2019.

\bibitem{Balasubramanian:2019wgd}
V.~Balasubramanian, M.~Decross, A.~Kar, and O.~Parrikar, ``{Quantum Complexity
  of Time Evolution with Chaotic Hamiltonians},'' {\em JHEP}, vol.~01, p.~134,
  2020.

\bibitem{Auzzi:2020idm}
R.~Auzzi, S.~Baiguera, G.~B. De~Luca, A.~Legramandi, G.~Nardelli, and
  N.~Zenoni, ``Geometry of quantum complexity,'' {\em Phys. Rev. D}, vol.~103,
  p.~106021, May 2021.

\bibitem{Balasubramanian:2021mxo}
V.~Balasubramanian, M.~DeCross, A.~Kar, Y.~C. Li, and O.~Parrikar,
  ``{Complexity growth in integrable and chaotic models},'' {\em JHEP},
  vol.~07, p.~011, 2021.

\bibitem{Magan:2018nmu}
J.~M. Mag\'an, ``{Black holes, complexity and quantum chaos},'' {\em JHEP},
  vol.~09, p.~043, 2018.

\bibitem{Jefferson_2017}
R.~Jefferson and R.~C. Myers, ``{Circuit complexity in quantum field theory},''
  {\em JHEP}, vol.~10, p.~107, 2017.

\bibitem{Chapman:2017rqy}
S.~Chapman, M.~P. Heller, H.~Marrochio, and F.~Pastawski, ``{Toward a
  Definition of Complexity for Quantum Field Theory States},'' {\em Phys. Rev.
  Lett.}, vol.~120, no.~12, p.~121602, 2018.

\bibitem{Khan:2018rzm}
R.~Khan, C.~Krishnan, and S.~Sharma, ``{Circuit Complexity in Fermionic Field
  Theory},'' {\em Phys. Rev. D}, vol.~98, no.~12, p.~126001, 2018.

\bibitem{Hackl_2018}
L.~Hackl and R.~C. Myers, ``{Circuit complexity for free fermions},'' {\em
  JHEP}, vol.~07, p.~139, 2018.

\bibitem{Chapman_2019}
S.~Chapman, J.~Eisert, L.~Hackl, M.~P. Heller, R.~Jefferson, H.~Marrochio, and
  R.~C. Myers, ``{Complexity and entanglement for thermofield double states},''
  {\em SciPost Phys.}, vol.~6, no.~3, p.~034, 2019.

\bibitem{Wu:2021pzg}
Q.-F. Wu, ``{Sectional curvatures distribution of complexity geometry},'' 8
  2021.

\bibitem{Note2}
The large $N$ limit we consider is similar to the vector large $N$ limit of
  $O(N)$ models of quantum field theories, where fields transform in the
  fundamental representation of the symmetry group, and the number of degrees
  of freedom $N$ is taken to infinity \cite {Klebanov:2018fzb}.

\bibitem{Note3}
We define an ideal fluid as being incompressible and inviscid.

\bibitem{arnold2008topological}
V.~Arnold and B.~Khesin, {\em Topological Methods in Hydrodynamics}.
\newblock Applied Mathematical Sciences, Springer New York, 2008.

\bibitem{Note4}
See Supplemental Material [URL inserted by the publisher] for more details.

\bibitem{FAIRLIE1989203}
D.~Fairlie, P.~Fletcher, and C.~Zachos, ``Trigonometric structure constants for
  new infinite-dimensional algebras,'' {\em Physics Letters B}, vol.~218,
  no.~2, pp.~203--206, 1989.

\bibitem{FAIRLIE1989101}
D.~B. Fairlie and C.~K. Zachos, ``{Infinite Dimensional Algebras, Sine Brackets
  and SU($\infty$)},'' {\em Phys. Lett. B}, vol.~224, pp.~101--107, 1989.

\bibitem{Fairlietrigonometric}
D.~B. Fairlie, P.~Fletcher, and C.~K. Zachos, ``{Infinite Dimensional Algebras
  and a Trigonometric Basis for the Classical Lie Algebras},'' {\em J. Math.
  Phys.}, vol.~31, p.~1088, 1990.

\bibitem{Patera}
J.~Patera and H.~Zassenhaus, ``The pauli matrices in n dimensions and finest
  gradings of simple lie algebras of type $a_{n-1}$,'' {\em Journal of
  Mathematical Physics}, vol.~29, no.~3, pp.~665--673, 1988.

\bibitem{Note5}
We show that the stream functions defined here are in one-to-one correspondence
  with the stream functions of hydrodynamics in section \hyperlink
  {Section3}{3}, see also \cite {Note4}.

\bibitem{PhysRevA.46.6417}
J.~S. Dowker and A.~Wolski, ``Finite model of two-dimensional ideal
  hydrodynamics,'' {\em Phys. Rev. A}, vol.~46, pp.~6417--6430, Nov 1992.

\bibitem{Note6}
In the context of holography, one is interested in target unitaries describing
  the time evolution of the system under a physical Hamiltonian $H$ up to a
  given time $t$, i.e. $U=e^{Ht}$. The parameter $s$ in the \protect \textit
  {control} Hamiltonian $\protect \mathcal {H}$ should not be confused with the
  time $t$ of the \protect \textit {physical} Hamiltonian $H$, as these are in
  general not equivalent. We assume the physical Hamiltonian implements
  all-to-all level transitions within the qudit \cite {Note4}.

\bibitem{AIF_1966__16_1_319_0}
V.~Arnold, ``Sur la g\'eom\'etrie diff\'erentielle des groupes de lie de
  dimension infinie et ses applications \`a l'hydrodynamique des fluides
  parfaits,'' {\em Annales de l'Institut Fourier}, vol.~16, no.~1,
  pp.~319--361, 1966.

\bibitem{Note7}
The restriction to $\protect \mathcal {O}(1)$ generators is justified by our
  choice of penalty factors as we elucidate in the following.

\bibitem{goldstein2002classical}
H.~Goldstein, C.~Poole, and J.~Safko, {\em Classical Mechanics}.
\newblock Addison Wesley, 2002.

\bibitem{Note8}
Here, $f_{\protect \textrm {target}}$ can be also taken to be time evolution of
  the fluid, assuming the physical Hamiltonian obeys similar properties as the
  aforementioned qudit Hamiltonian \cite {Note4}.

\bibitem{Note9}
See eq.~(S20) in \cite {Note4}.

\bibitem{Note10}
Our definition of the discrete Laplacian is the canonical one but it is
  non-unique, see e.g. \cite {PhysRevA.46.6417}.

\bibitem{Note11}
See however \cite {Magan:2018nmu} where polynomial penalties were first
  suggested.

\bibitem{Note12}
The Pauli basis is used to decompose many-qubit gates into tensor products of
  the standard Pauli matrices and the identity. A gate is of weight $k$ (or
  \protect \textit {$k$-local}) if its tensor product contains up to $k$ Pauli
  matrices.

\bibitem{agrachev2019}
A.~Agrachev, D.~Barilari, and U.~Boscain, {\em A Comprehensive Introduction to
  Sub-Riemannian Geometry}.
\newblock Cambridge Studies in Advanced Mathematics, Cambridge University
  Press, 2019.

\bibitem{Brown:2021rmz}
A.~R. Brown, M.~H. Freedman, H.~W. Lin, and L.~Susskind, ``{Effective Geometry,
  Complexity, and Universality},'' 11 2021.

\bibitem{Rashevskii1938}
P.~K. Rashevsky, ``Any two points of a totally nonholonomic space may be
  connected by an admissible line,'' {\em Uch. Zap. Ped. Inst. im. Liebknechta,
  Ser. Phys. Math.}, vol.~2, p.~83–94, 1938.
\newblock (in Russian).

\bibitem{Chow1940}
W.-L. Chow, ``Ueber systeme von linearen partiellen differentialgleichungen
  erster ordnung,'' {\em Mathematische Annalen}, vol.~117-117, pp.~98--105,
  Dec. 1940.

\bibitem{KhajehSalehani2014ControllabilityOI}
M.~K. Salehani and I.~Markina, ``Controllability on infinite-dimensional
  manifolds: A chow–rashevsky theorem,'' {\em Acta Applicandae Mathematicae},
  vol.~134, pp.~229--246, 2014.

\bibitem{Note13}
The applicability of this theorem to our setup is discussed in detail in the
  SM.

\bibitem{Arnold2014}
V.~I. Arnold, {\em Exponential scattering of trajectories and its
  hydrodynamical applications}, pp.~419--427.
\newblock Berlin, Heidelberg: Springer Berlin Heidelberg, 2014.

\bibitem{Lukatskii1981}
A.~M. Lukatskii, ``On the curvature of the group of measure-preserving
  diffeomorphisms of an n-dimensional torus,'' {\em Russian Mathematical
  Surveys}, vol.~36, pp.~179--180, Apr. 1981.

\bibitem{Note14}
A similar behavior of negative curvature only in a subsector of the manifold of
  unitaries, built out of the Pauli basis, was found in the context of operator
  size complexity in \cite {Wu:2021pzg}.

\bibitem{MR0224110}
D.~V. Anosov, ``Geodesic flows on closed {R}iemannian manifolds of negative
  curvature,'' {\em Trudy Mat. Inst. Steklov.}, vol.~90, p.~209, 1967.

\bibitem{Milnor}
J.~Milnor, ``Curvatures of left invariant metrics on lie groups,'' {\em
  Advances in Mathematics}, vol.~21, pp.~293--329, 9 1976.

\bibitem{bishop2013riemannian}
R.~L. {Bishop}, ``{Riemannian Geometry},'' {\em arXiv e-prints},
  p.~arXiv:1303.5390, Mar. 2013.

\bibitem{10.2307/2161966}
G.~Misiołek, ``Conjugate points in
  \texorpdfstring{$\mathcal{D}_{\mu}(\mathds{T}^2)$}{Du(T2)},'' {\em
  Proceedings of the American Mathematical Society}, vol.~124, no.~3,
  pp.~977--982, 1996.

\bibitem{Klebanov:2018fzb}
I.~R. Klebanov, F.~Popov, and G.~Tarnopolsky, ``{TASI Lectures on Large $N$
  Tensor Models},'' {\em PoS}, vol.~TASI2017, p.~004, 2018.

\bibitem{Gottesman:1998se}
D.~Gottesman, ``{Fault tolerant quantum computation with higher dimensional
  systems},'' {\em Chaos Solitons Fractals}, vol.~10, pp.~1749--1758, 1999.

\bibitem{doi:10.1137/S0097539799359385}
D.~Aharonov and M.~Ben-Or, ``Fault-tolerant quantum computation with constant
  error rate,'' {\em SIAM Journal on Computing}, vol.~38, no.~4,
  pp.~1207--1282, 2008.

\bibitem{PhysRevLett.113.230501}
E.~T. Campbell, ``Enhanced fault-tolerant quantum computing in $d$-level
  systems,'' {\em Phys. Rev. Lett.}, vol.~113, p.~230501, Dec 2014.

\bibitem{connes1995noncommutative}
A.~Connes, {\em Noncommutative Geometry}.
\newblock Elsevier Science, 1995.

\bibitem{Hoppe1992}
J.~Hoppe, {\em Lectures on Integrable Systems}.
\newblock Springer Berlin Heidelberg, 1992.

\bibitem{Khesin_2004}
B.~Khesin, A.~Levin, and M.~Olshanetsky, ``Bihamiltonian structures and
  quadratic algebras in hydrodynamics and on non-commutative torus,'' {\em
  Communications in Mathematical Physics}, vol.~250, p.~581–612, Aug 2004.

\bibitem{Note15}
We adopt the convention of the mathematical literature and take anti-Hermitian
  instead of Hermitian matrices.

\bibitem{Note16}
By \enquote {compatibility} we mean that the structure of the distribution
  remains the same and subsequent distributions are a natural extension of
  previous ones. This allows for the systematic generation of the distribution.

\bibitem{Note17}
Trivially flat directions are those given by commuting generators.

\bibitem{Note18}
The Chow-Rashevskii theorem ensures this is possible.

\bibitem{Note19}
Here we have assumed all the hard directions are degenerate for simplicity. The
  more general case is treated in \cite {Brown:2021rmz}.

\bibitem{susskind2014entanglementisnotenough}
L.~Susskind, ``{Entanglement is not enough},'' {\em Fortsch. Phys.}, vol.~64,
  pp.~49--71, 2016.

\bibitem{Haferkamp:2021uxo}
J.~Haferkamp, P.~Faist, N.~B.~T. Kothakonda, J.~Eisert, and N.~Y. Halpern,
  ``{Linear growth of quantum circuit complexity},'' 6 2021.

\end{thebibliography}

\end{document}